\documentclass[traditabstract]{aa}  

\usepackage{graphicx}
\usepackage{txfonts}
\usepackage[usenames,dvipsnames]{color} 
\usepackage{longtable,lscape}
\usepackage{natbib}
\bibpunct{(}{)}{;}{a}{}{,} 
\newcommand{\Msun}{\mbox{ $\mathrm{M}_{\odot}$}}

\begin{document}

   \title{Binary progenitor models of type IIb supernovae}
   \author{J.S.W. Claeys
          \inst{1,2}
          \and
          S.E. de Mink\inst{1,3,4}\thanks{Hubble fellow.}
          \and
          O.R. Pols\inst{1}
          \and
          J.J. Eldridge\inst{5}
          \and
          M. Baes\inst{2}
          }
   \institute{ Sterrekundig Insituut, Universiteit Utrecht, PO Box 800000, 3508 TA Utrecht, The Netherlands
\and   
   Sterrenkundig Observatorium, Universiteit Gent, Krijgslaan 281-S9, B-9000 Gent, Belgium
\and   
   Argelander Institute for Astronomy, University of Bonn, Auf dem Huegel 71, D-53121 Bonn, Germany;
\and  
   Space Telescope Science Institute, 3700 San Martin Drive, Baltimore, MD 21218, USA. Hubble Fellow;
\and  
   Institute of Astronomy, University of Cambridge, Madingley Road, Cambridge CB3 0HA\\
                         \email{J.Claeys@uu.nl, S.E.deMink@gmail.com, O.R.Pols@uu.nl}
             }

   \date{Received 16 July 2010 / Accepted 30 January 2011}

\abstract{
Massive stars that lose their hydrogen-rich envelope down to a few tenths of a solar mass explode as extended type IIb supernovae, an intriguing subtype that links the hydrogen-rich type II supernovae with the hydrogen-poor type Ib and Ic.  The progenitors may be very massive single stars that lose their envelope due to their stellar wind, but mass stripping due to interaction with a companion star in a binary system is currently considered to be the dominant formation channel. 

Anticipating the upcoming automated transient surveys, we computed an extensive grid of binary models with the Eggleton binary evolution code. We identify the limited range of initial orbital periods and mass ratios required to produce type IIb binary progenitors. The rate we predict from our standard models, which assume conservative mass transfer, is about six times smaller than the current rate indicated by observations. It is larger but still comparable to the rate expected from massive single stars.  We evaluate extensively the effect of various assumptions such as the adopted accretion efficiency, the binary fraction and distributions for the initial binary parameters. To recover the observed rate we must generously allow for uncertainties and consider low accretion efficiencies in combination with limited angular momentum loss from the system. 

Motivated by the claims of detection and non-detection of companions for a few IIb supernovae, we investigate the properties of the secondary star at the moment of explosion.  We identify three cases: (1) the companion is predicted to appear as a hot O star in about 90\% of the cases, as a result of mass accretion during its main sequence evolution, (2) the companion becomes an over-luminous B star in about 3\% of the cases, if mass accretion occurred while crossing the Hertzsprung gap or (3) in systems with very similar initial masses the companion will appear as a K supergiant.  The second case, which applies to the well-studied case of SN 1993J and possibly to SN 2001ig, is the least common case and requires that the companion very efficiently accretes the transferred material -- in contrast to what is required to recover the overall IIb rate. We note that relative rates quoted above depend on the assumed efficiency of semi-convective mixing: for inefficient semi-convection the presence of blue supergiant companions is expected to be more common, occurring in up to about 40\% of the cases.

Our study demonstrates that type IIb supernovae have the potential to teach us about the physics of binary interaction and about stellar processes such as internal mixing and possibly stellar-wind mass loss.  The fast increasing number of type IIb detections from automated surveys may lead to more solid constraints on these model uncertainties in the near future.
 }

\keywords
{Stars: evolution - Binaries: general - Supernovae: general - Supernova: individual: SN 1987K, SN 1993J, SN 1996cb, SN 2000H, SN 2001ig, SN 2001gd, SN 2003bg, SN2008ax, Cas A}

\maketitle

\section{Introduction}
Core collapse supernovae are the bright explosions marking the end of the lives of massive stars.  
Their light curves and spectral signatures come in a large variety of types and yield information about the structure and chemical composition of the progenitor star and it surroundings. 
Type II supernovae -- characterised by strong hydrogen lines -- are associated with massive stars that are still surrounded by their hydrogen-rich envelope at the time of explosion, whereas type Ib and Ic supernovae -- in which no signature of hydrogen is found -- are thought to result from massive stars that have lost their entire hydrogen-rich envelope.   

Type IIb supernovae constitute an intriguing intermediate case. Initially they show clear evidence for hydrogen, but later the hydrogen lines become weak or absent in the spectra.  Two famous examples of this subtype are SN 1993J  \cite[]{Ripero93} and Cassiopeia A, which was recently classified using the scattered light echo \citep{Krause08}. 
The typical light curve of a type IIb supernova, such as SN 1993J, is characterized by two peaks. The first maximum is associated with shock heating of the hydrogen-rich envelope, resembling a type II supernova.  The second maximum is caused by the radioactive decay of nickel \cite[]{Benson94}. 
These characteristics can be explained assuming that the progenitor star had an extended low-mass hydrogen envelope at the time of explosion, with between 0.1 and 0.5 M$_\odot$ of hydrogen \citep{Podslad93, Woosley94, Elmhamdi06}.  Progenitors with smaller hydrogen envelope masses are compact, but can be classified as type IIb supernovae if hydrogen lines are detected shortly after the explosion, e.g. \citet{Chevalier+Soderberg2010}.  In this paper we focus on the progenitors of extended type IIb supernovae.

Massive single stars with masses of about 30~\Msun~or higher can lose their envelope as a result of a stellar wind.  Alternatively, massive stars can be stripped of their envelope due to interaction with a companion star in a binary system. Quickly after the discovery of SN 1993J many authors realized that the single star scenario requires precise fine tuning of the initial mass of the star in order to have a very low-mass envelope left at the time of the explosion \citep[][and references therein]{Podslad93,Woosley94}.  
According to these authors,  mass stripping from an evolved red supergiant in a binary system naturally leads to a hydrogen envelope containing a few times 0.1~\Msun~at the moment of explosion.  
More compact binaries, where the primary star is stripped in a less evolved stage, result in smaller hydrogen envelopes  and may lead to type IIb supernovae of the compact category \cite[e.g.][]{YoonWL10} or type Ib supernovae. 

 \citet{Nomoto93} proposes a third scenario which involves a common envelope phase and finally a merger of the core of the primary star and its companion.  Energy to eject the envelope is extracted from the orbital energy, but a small layer of hydrogen may remain on the surface of the merger product until the moment of explosion as a supernova type IIb.  This scenario was recently considered by \citet{Young06} for Cassiopeia A, in which no evidence for a companion star was found, despite numerous attempts.

A decade after the detection of 1993J,  the interest for type IIb supernova revived when \citet{Maund04} found evidence for the presence of a blue supergiant companion at the location where 1993J had just faded away. These findings nicely fitted the predictions for the companion star in the models by \citet{Podsiadlowski92, Podslad93}. 
\citet{Stancliffe09} performed the first systematic study of binary progenitor models that matched the properties of the progenitor of 1993J and its companion. They computed a grid of binary models varying the initial primary masses and initial orbital periods. Even though they find a suitable progenitor model, they emphasize that it proves to be extremely difficult to explain the properties of the blue supergiant companion star. 

At the time of writing about 69 supernovae have been classified as IIb supernova\footnote{http://www.cfa.harvard.edu/iau/lists/Supernovae.html}.  This number will increase quickly thanks to upcoming systematic wide-field surveys of transient events, such as the Palomar Transient Factory \citep[e.g.][]{Arcavi+10}, Pan-STARRS \citep{PanSTARRS2002} and the Large Synoptic Survey Telescope \citep{LSST2008}. About 9 have been studied well enough such that constraints can be derived on the progenitors and in some cases on possible companion stars.  

These observational developments and the questions raised by \citet{Stancliffe09} motivated us to further investigate binary progenitor models for type IIb supernova. In contrast with previous work we will not just discuss the case of SN 1993J.  Instead we discuss the properties of the progenitor stars and their companions in general. We extend the work of  \citet{Stancliffe09} by exploring the parameter space of binary progenitor models in detail, in particular the dependence on the initial mass ratio.  Furthermore we discuss the properties of the companion star at the time of explosion. 

In section~\ref{sec:obs} we review observed type IIb supernovae and the observed rates. In section~\ref{sec:model} we outline the assumptions in our binary evolution code.  In section~\ref{sec:single}  we briefly discuss single-star progenitor models.  Section~\ref{sec:binary} we discuss our progenitor models and in section~\ref{sec:rates} the rate of IIb supernovae predicted from our models. Section~\ref{sec:disc} contains a  discussion, conclusion and outlook.

\section{Observed type IIb Supernovae\label{sec:obs}}
In recent years about 69 supernovae of type IIb have been detected$^1$. These observations enabled the first estimates for the rate of type IIb supernovae. The determination of the rate is not straightforward due to observational biases and selection effects. For example, if the supernova is observed too late it will be classified as a type Ib supernova instead of type IIb \citep[see also][]{Maurer+2010arXiv}. \cite{VLF05} and \cite{LWVetal07} derive that $3.2 \pm 1.0\%$ and $1.5 \pm 1.5\%$ of all core collapse SNe (CCSNe) are of type IIb. They based their estimates on the discoveries by the  Lick Observatory Supernova Search (LOSS), respectively using a 140 and 30 Mpc distance-limited sample. \cite{VLF05}  also specified that the distribution of SNe IIb and SNe II does not depend on the morphological type of the host galaxy. \citet{Smartt09, Smartt09_review} evaluated different determinations of the rate and the effect of observational biases.  They estimate the rate to be $5.4 \pm 2.7 \%$ based on observations covering 10.5 years within 28 Mpc.  However, there are only 5 IIb supernovae in their sample, of which at least two are compact IIb supernovae. Preliminary results of the Palomar Transient Factory survey point towards a rate of $3.6\pm 2.5 \%$ in giant hosts and a larger fraction in dwarf galaxies, $20\pm11\%$ \cite[]{Arcavi+10}. 

While this study was near completion \citet{Smith+2010arXiv} presented the results of a homogeneous volume limited survey within 60 Mpc which includes 80 CCSNe. They find a significantly higher fraction of type IIb supernova, $10.6^{+3.6}_{-3.1}\%$, with respect to previous studies. They attribute the difference to their more complete photometric and spectroscopic follow-up observations used to classify type II supernovae into different subtypes. It is unclear how many of the type IIb supernovae quoted in these studies are extended instead of compact.

For the purpose of comparison with our model predictions we will use the rate quoted by \cite{Smartt09} after taking out the two compact Type IIb, resulting in a fraction of extended type IIb of about 3\% with respect to all core collapse supernovae. We emphasize that this number is still very uncertain. More reliable determinations of the rate of extended type IIb supernovae are expected in the near future from the current and upcoming automated transient surveys."

In the follow paragraphs we discuss some observed supernovae type IIb and discuss their observed characteristics.

\subsection{Individual type IIb supernovae}

\subsubsection{\object{SN 1987K}}
SN 1987K was the first supernova observed with the characteristics of a type IIb supernova, namely the transition from a supernova type II to type Ib \cite[]{Filippenko88}, but it was only later that it was defined a SN of type IIb. \cite{Filippenko88} already suggested that these characteristics could be due to mass loss of a massive star, with an initial mass between 20 and 25 M$_\odot$ or a combination of mass transfer and winds of a less massive star, between 8 and 20 M$_\odot$.

\subsubsection{\object{SN 1993J} \label{sec:1993J}}
SN 1993J was the first supernova classified as type IIb. This supernova has been well studied because the progenitor was detected and recognized as a star with a spectral class K0Ia \cite[]{Filip93}. Its binary companion has been observed recently as an early B-supergiant \citep[with best estimate a B2Ia star, ][]{Maund09}. The light curve of the supernova can be explained if the star had an amount of hydrogen in the envelope of about 0.1--0.5~M$_\odot$ at time of explosion \cite[]{Woosley94, Filip93}. \cite{Maund04} determined the luminosity and effective temperature of the progenitor of the supernova and its companion, namely $\log L/L_\odot = 5.1 \pm 0.3$ and $\log T_{\rm eff}/{\rm K} = 3.63 \pm 0.05$ for the progenitor and $\log L/L_\odot = 5 \pm 0.3$ and $\log T_{\rm eff}/{\rm K} = 4.3 \pm 0.1$ for the companion. The ejecta mass was determined, using the width of the second peak of the light curve, to be about 4 $M_\odot$ \cite[]{Shige94}. Radio and X-ray observations showed evidence for a mass loss rate of about $4 \times 10^{-5}$ \Msun~yr$^{-1}$ at the time of explosion \cite[]{Fransson96}.

\subsubsection{\object{SN 1996cb}}
Supernova 1996cb was also determined to be of type IIb \cite[]{Qiu99}. The light curve of this supernova was not observed until several days after the explosion. Instead of the two peaks that are typical for type IIb light curves, the light curve consists of a short-term plateau phase similar to a type II supernova. These differences with SN 1993J arise from a more massive hydrogen envelope of SN 1996cb \cite[]{Qiu99}.

\subsubsection{\object{SN 2000H}}
\cite{Benetti00} classified SN 2000H a type IIb supernova because of the hydrogen lines observed in the spectrum. However, these hydrogen lines were not obvious and the light curve showed more resemblance with a SN Ib \cite[]{Branch02}. Therefore, this supernova has been classified by other authors as a type Ib supernova \citep[]{Branch02, Elmhamdi06}. The amount of hydrogen was estimated to be 0.08 M$_\odot$ \cite[]{Elmhamdi06}.

\subsubsection{\object{SN 2001gd}}
The observations of radio light curves give an estimate of the mass loss rate of the progenitor system of the supernova. The radio light curve of the type IIb SN 2001gd indicates a mass loss rate of about 2--12 $\times$ $10^{-5}$ M$_\odot$yr$^{-1}$ \citep[]{Stockdale03, Perez-Torres05}, which is in between the rates for typical type II and type Ib SNe.

\subsubsection{\object{SN 2001ig}}
Another supernova of type IIb, SN 2001ig, was observed in 2001 in NGC 7424 and shows a spectral evolution similar to that of SN 1993J \cite[]{Ryder06}. Evidence was found for a star of spectral type late-B through late-F at the location of SN 2001ig, a possible companion of the progenitor of SN 2001ig \cite[]{Ryder06}.

\subsubsection{\object{SN 2003bg}}
SN 2003bg evolved from a type Ic supernova to a hydrogen-rich type IIb, to a hydrogen-poor type Ibc \cite[]{Soderberg06}. It was observed as a broad-lined type IIb supernova and proclaimed to be 'the first type IIb hypernova' \citep[]{Mazzali09, Hamuy09}. The broadness of the lines indicates a high progenitor mass \cite[]{Hamuy09}. The light curve and spectral evolution indicate the presence of a thin layer of hydrogen at time of explosion, $\approx 0.05$\Msun~\cite[]{Mazzali09}. The velocity of the ejecta resembles more closely the velocity of SNe type Ib than type II \cite[]{Soderberg06}. This implies a compact Wolf-rayet progenitor, with a progenitor mass between 20 and 25~\Msun. \cite{Soderberg06} conclude that this event is an intermediate case between SNe type IIb and type Ib.

\subsubsection{\object{SN 2008ax}}
The light curve of SN 2008ax shows some differences with the light curve of SN 1993J, namely the lack of the first peak  and it has slightly bluer colors \cite[]{Pastor08}. These features can be explained by a less massive hydrogen envelope at time of explosion, less than a few $\times$ 0.1 M$_\odot$, in comparison with the progenitor of SN 1993J \cite[]{Crocket08}.

\subsubsection{\object{Cas A}}
Cas A is the supernova remnant of a star that exploded about 350 years ago \cite[]{Thorstensen01}. A light echo from this explosion \cite[]{Krause08} shows evidence that it was a supernova of type IIb. Direct methods to determine the mass of the progenitor star are difficult, but the ejecta mass was calculated to be 2--4 M$_\odot$ and the remnant would be expected to be a neutron star with a mass between 1.5 and 2.2 M$_\odot$ \cite[]{Young06}. This sets the mass of the star at time of its explosion at about 4--6 M$_\odot$. There is no direct evidence as to wether this supernova was of the compact or extended type IIb, but the possibilty that the progenitor was a red supergiant is left open. For this supernova single and binary progenitor models were calculated \cite[]{Young06}. The single star models indicated fine-tuning of the stellar wind is necessary to evolve to the specific characteristics of the supernova remnant Cas A \cite[]{Young06}. Besides, there is evidence that the progenitor could only have had a very short-lived Wolf-rayet phase, which is difficult to explain with single stars \citep[]{Schure08, VanVeelen09}.  There is no evidence for a companion star. Therefore a common envelope scenario was proposed, in which the two stars merge into a single star before explosion. Observations show tentative evidence for this scenario \cite[]{Krause08}, such as the asymmetric distribution of the quasi-stationary flocculi near Cas A, which could arise from the loss of a common envelope.\\
\\The observations put constraints on the general properties of a SN type IIb: the explosion of a supergiant with a hydrogen envelope mass between 0.1 and 0.5 M$_\odot$. We consider the lower limit of the mass of the hydrogen envelope to be 0.1~\Msun~rather than 0.01~\Msun, the lower limit proposed by \cite{Chevalier+Soderberg2010}. The explosion of a star with a hydrogen envelope mass smaller than 0.1~\Msun~will exhibit hydrogen lines in its spectrum, but only in the early phases of the supernova. In addition the light curve resembles the typical light curve of a SN type Ib. Therefore such a supernova will more likely be defined as a supernova type Ib or a transitional type between IIb and Ib (see examples above, e.g. SN 2003bg and SN 2000H). \cite{Elmhamdi06} places the upper limit of the mass of the hydrogen envelope of the progenitor of a type Ib supernova at 0.1~\Msun. 

A binary progenitor is confirmed or considered likely in some cases, and the secondary has been detected as a blue supergiant in possibly two cases. However, no general constraints on the secondary can be set.

\section[]{Stellar evolution calculations \label{sec:model}}

We use a version of the binary evolution code STARS originally developed by \citet{Eggleton71} and later updated and described by various authors \citep[e.g.][]{Pols+95, EggletonsBook06, Glebbeek+08}. The code is fully implicit and solves the equations for the structure and 
composition of the star simultaneously.  It employs an adaptive non-Lagrangian mesh that places mesh points in regions of the star where higher resolution is required.  This allows us to evolve stars with a reasonable accuracy using as few as 200 mesh points. The code therefore is fast and suitable to compute the large numbers of models needed to investigate wide initial parameter space of binary systems \citep[e.g.][]{DeMink+07}. 

We use nuclear reaction rates from \citet{Caughlan+1985} and \citet{Caughlan+Fowler1988} and opacities from \citet{Rogers+Iglesias1992} and \citet{Alexander+Ferguson1994}. The assumed heavy-element composition is scaled to solar abundances \citep{Anders+Grevesse1989}.

Convection is implemented using a diffusion approximation \citep{Eggleton72} of the mixing-length theory
\citep{Boehm-Vitense1958}, assuming a mixing length of 2.0 pressure scale heights. We use the Schwarzschild criterion to determine the boundaries of the convective regions. Convective overshooting is taken into account using the prescription of \cite{Schroder97} with an overshooting parameter of $\delta_{\rm ov}$ = 0.12, which was calibrated against accurate stellar data eclipsing binaries \citep{Pols+1997}.  In terms of the pressure scale height, as the overshooting parameter is commonly defined in other stellar evolution codes, this value approximately compares to $\alpha_{\rm ov} \approx 0.25$. 

Mass loss in the form of a stellar wind is taken into account adopting the prescription by \cite{deJager1988}. Although for O and B stars this prescription has been superseded by more recent mass-loss determinations, this is not the case for the red supergiant region of the H-R diagram where the only significant mass loss in our binary models occurs. For computational reasons we ignore stellar-wind mass loss from the less massive companion star and we ignore any possible accretion from the stellar wind of the primary.

To compute the evolution of interacting binaries we evolve the two stars quasi-simultaneously. First we follow the primary for several steps and store the changes of the masses of both stars and of the orbit. Afterwards the secondary is evolved applying these mass changes until its age reaches that of the primary.
When the primary star expands beyond its Roche lobe we compute the mass-transfer rate as a function of the difference in potential between the Roche-lobe surface and the stellar surface.  The mass flux of each mesh point beyond the Roche lobe is given by
\begin{equation}
\label{functMtr}
\frac{{\rm d}\dot{M}}{{\rm d}m} = -C \times \frac{\sqrt{2\phi_{\rm s}}}{r},
\end{equation}
as in \cite{DeMink+07}, where $m$ and $r$ denote the mass and radius coordinate of the mesh point and  $\phi_{\rm s}$ the difference in potential with respect to the Roche-lobe surface.  $C$ is a proportionality constant which we set to 10$^{-2}$ for numerical convenience. The mass transfer rate is then given by the integral of Eq. (\ref{functMtr}) over all mesh points outside the Roche radius. 
As a result of the implementation of mass transfer in our code we implicitly assume that the entropy and composition of the accreted material is equal to the entropy and composition of material at the surface of the accreting star \citep[e.g.][]{Pols1994}.
Tidal effects are not taken into account in this work.

To investigate the effects of non-conservative mass transfer, i.e. mass and angular momentum loss from the system during Roche-lobe overflow, we assume that a constant fraction $\beta$ of the transferred mass is accreted by the companion.  We adopt different values: $\beta =1$ (conservative mass transfer) and  $\beta= 0.75, 0.5, 0.25$ (non-conservative mass transfer). The specific angular momentum of the mass lost from the system is assumed to be the specific angular momentum of the orbit of the accreting star. 

Using this code, we calculated a grid of binary systems for an initial primary mass of 15 M$_\odot$. We varied the initial orbital period  between 800 and 2100 days in steps of 100 days and we varied the initial secondary mass between about 10 and 15~\Msun~in steps of 0.1 M$_\odot$.  For systems with similar masses we increased the resolution by varying the secondary mass in steps of 0.01~\Msun. In addition we computed several systems with different primary masses. 
 
The stars are evolved until the onset of carbon burning. The star will explode shortly after this point, about 2.6 $\times$ 10$^3$ years for a star of 15 M$_\odot$ \cite[]{Eid04}.  
To determine the masses of the stars and in particular of their envelopes at the time of explosion we extrapolate the mass-loss rate by stellar winds and Roche-lobe overflow to determine the masses at time of explosion.
When both stars fill their Roche lobe at the same time and a contact binary is formed, we end our simulation.  In this case the binary system will probably evolve into a common envelope.
 
We note that whether or not a contact system forms is sensitive to some of our assumptions regarding mass transfer. The chosen value of $C$ in eq.~(\ref{functMtr}) leads to a maximum mass-transfer rate that is lower than the self-regulated rate on the thermal timescale of the donor star. If we choose a larger value of $C$ the mass-transfer rate increases, which affects the response of the secondary star. 
Since mass transfer is faster than the thermal timescale of the secondary, more rapid accretion results in a stronger radius expansion of the secondary and a higher likelihood of forming a contact system. 
On the other hand, the transferred material comes from the surface of a red supergiant which has a much smaller specific entropy than the hot surface of the secondary. Although the gas may undergo additional heating during accretion, this is probably not sufficient to make its entropy equal to the surface entropy of the secondary as we implicitly assume. Taking this into account properly, which is very difficult, would likely result in less radius expansion of the secondary. These two simplifying assumptions thus have opposite effects, making the true boundary between systems that do and do not come into contact hard to predict from our models (see Section~\ref{sec:parameterspace} and Section~\ref{sec:disc}).

\begin{table} 
\caption{Properties of single stellar models for different initial masses $M_{\rm i}$.}
\label{TS}
\begin{center}
\begin{tabular}{|c|c|c|c|c|c|}
\hline \hline   
\begin{tabular}{c}
$M_{\rm i}$ \\($M_\odot$)\\ 
\end{tabular} 
& \begin{tabular}{c}
$M_{\rm f}$ \\($M_\odot$)\\ 
\end{tabular} 
& \begin{tabular}{c} 
$M_{\rm H}$\\($M_\odot$)\\
\end{tabular} 
& $\log \left(\frac{L}{L_\odot}\right)$ & 
\begin{tabular}{c}
$T_{\rm eff}$\\($10^3$K)\\ 
\end{tabular} & 
SN type
\\
\hline
32 & 13.94 & 1.014 & 5.56 & 3.6 & IIL\\
32.5 & 12.89 & 0.505 & 5.57 & 3.8 & IIb\\
33 & 11.81 & 0.080 & 5.58 & 4.3 & Ib\\
\hline
\end{tabular}
\end{center}
\tablefoot{The final mass M$_{\rm f}$ and the amount of hydrogen in the envelope M$_{\rm H}$ are given at the time of explosion, together with the luminosity log(L/L$_\odot$) and effective temperature T$_{\rm eff}$ at the onset of central carbon burning.}
\end{table}

\section{Single star progenitors \label{sec:single}}

At the time of explosion the progenitor of a type IIb supernova is surrounded by a low-mass hydrogen envelope.  In massive single stars the envelope can be removed by the stellar wind. \citet{Heger+2003} find that at solar metallicity the range for type IIb and IIL supernovae combined ranges from roughly 26--34~\Msun. Type IIL supernovae show a linear decline in the light curve, but without transition to a type Ib, which requires a hydrogen-rich envelope of less than about 2~\Msun~at the time of explosion. \citet{Eldridge+Tout2004} map single-star progenitors of different types of supernovae as a function of metallicity.   At solar metallicity they find that IIL and IIb supernova progenitors combined should have initial masses between approximately 25 and 30~\Msun. \citet{Georgy+2009} discuss single star SN progenitors from rotating models at different metallicities.  Although they do not discuss type IIb or IIL in particular, they find that the transition from type II to Ib should occur around 25~\Msun. \citet{Perez+2009} investigate possible single star progenitors for Cas A and need a progenitor of approximately 30~\Msun. In particular, a single star scenario has been proposed for hypernova SN 2003bg to explain the high mass. On the other hand a single star scenario cannot explain the characteristics of SN 1993J (see Sec. \ref{sec:1993J}). Consequently, the single star scenario cannot explain all observed type IIb supernovae.

The determination of the rate from single stars is sensitive to uncertainties in the stellar wind mass loss rates. Especially with respect to stellar winds from massive red giants our understanding is limited.  \cite{YoonCantiello10} speculate that a superwind driven by pulsational instabilities may drive a strong mass loss, bringing the minimum mass for type IIb supernovae down to about 20~\Msun. 

Single stars with a mass above $\sim$25~\Msun~are believed to produce only faint supernovae \cite[]{Fryer99}. Consequently, these type IIb SNe will appear different than type IIb SNe formed by binary stars. Nevertheless, in the correct mass range, single stars can explode as type IIb SNe and therefore it is reasonable to compare the expected rate from binary and single-star progenitors. To be able to do this comparison we computed single stellar models with the same input physics as the binary models we discuss in the next section. We find that the initial mass range for single stars resulting in type IIb progenitors should be within 32.5--33~\Msun.  In Table~\ref{TS} we list the final mass and and the amount of hydrogen left in the envelope near this range of initial masses.  Assuming a Kroupa initial mass function \cite[]{Kroupa01} we estimate that about 0.3\% of all single stars more massive than 8 solar masses are within this mass range and will end their lives as a type IIb supernova.

\begin{figure}
\includegraphics[width=6cm,angle=270]{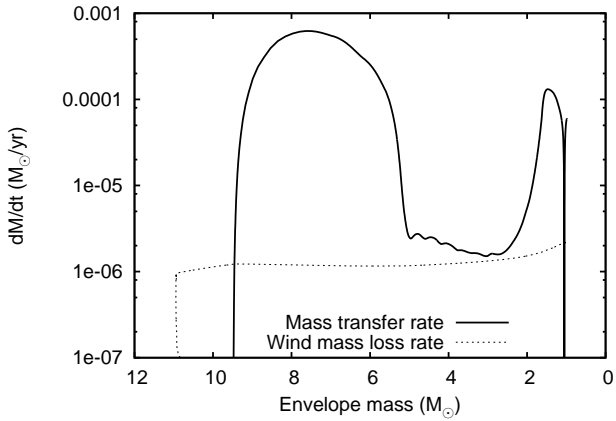}
\caption{Wind mass loss and mass transfer rate of as a function of the remaining envelope mass for the most massive star of binary system with initially 15+14.35 M$_\odot$, an initial orbital period of 1500 days. For details see Section~\ref{sec:binary}.}
\label{MtrMass}
\end{figure}

\section{Binary star progenitors \label{sec:binary}}

\begin{table*} 
\caption{Properties of binary stellar models for an initial primary mass of 15~\Msun, conservative mass transfer and with variation of the initial  secondary mass M$_{\rm 2i}$ and initial orbital period P$_{\rm orb}$.}
\label{TC1}
\begin{center}
\begin{tabular}{|c|c|c|c|c|c|c|c|c|c|}
\hline \hline   
$N^0$ & 
\begin{tabular}{c}
$M_{2i}$ \\($M_\odot$)\\ 
\end{tabular} 
& \begin{tabular}{c}
$M_{1f}$ \\($M_\odot$)\\ 
\end{tabular} 
& \begin{tabular}{c}
$M_{2f}$ \\($M_\odot$)\\ 
\end{tabular} 
& \begin{tabular}{c} 
$P_{\rm orb}$ \\ (days) \\
\end{tabular} 
& \begin{tabular}{c} 
$M_{H1}$\\($M_\odot$)\\
\end{tabular} 
& log$\displaystyle\left(\frac{L_1}{L_\odot}\right)$ & 
\begin{tabular}{c}
$T_{\rm eff,1}$\\($10^3$K)\\ 
\end{tabular} 
& log$\displaystyle\left(\frac{L_2}{L_\odot}\right)$ & 
\begin{tabular}{c}
$T_{\rm eff,2}$ \\($10^3$K)\\
\end{tabular} \\
\hline
$\#$1 & 14 & 5.93 & 20.92 & 1500 & 0.354 & 5.07 & 3.48 & 4.95 & 31.4\\
$\#$2 & 14.35 & 5.98 & 21.22 & 1500 & 0.370 & 5.07 & 3.47 & 5.06 & 17.9\\
$\#$3 & 14.55 & 6.01 & 21.40 & 1500 & 0.390 & 5.07 & 3.45 & 4.85 & 3.82\\
$\#$4 & 14.95 & 6.02 & 21.79 & 1400 & 0.401 & 5.07 & 3.51 & 5.07 & 3.63\\
\hline
\end{tabular}
\end{center}
\tablefoot{The final mass of the primary and secondary, M$_{\rm f1}$ and M$_{\rm f2}$, and the amount of hydrogen in the envelope $M_{\rm H}$ are given at the time of explosion, together with the luminosity log(L/L$_\odot$) and effective temperature T$_{\rm eff}$ of the primary and secondary at the onset of central carbon burning in the primary (except for model \#4, where L$_2$ and T$_{\rm eff,2}$ refer to the last computed model of the secondary).}
\end{table*}

\begin{table*} 
\caption{Expected broad-band magnitudes of the three different possible evolutionary paths of the companion star, with an intial mass M$_{2i}$.}
\label{TUV}
\begin{center}
\begin{tabular}{|c|c|c|c|c|c|c|c|c|c|}
\hline \hline   
$N^0$ & 
\begin{tabular}{c}
$\rm M_{2i}$ \\($\rm M_\odot$)\\ 
\end{tabular} 
& \begin{tabular}{c}
M(U)\\ 
\end{tabular} 
& \begin{tabular}{c}
M(B)\\ 
\end{tabular} 
& \begin{tabular}{c} 
M(V)\\
\end{tabular} 
& \begin{tabular}{c} 
M(R)\\
\end{tabular} 
& M(I) & 
\begin{tabular}{c}
M(J)\\ 
\end{tabular} 
& M(H) & 
\begin{tabular}{c}
M(K)\\
\end{tabular} \\
\hline
$\#$1 & 14 & -5.89 & -4.99 & -4.65 & -4.66 & -4.37 & -4.02 & -3.91 & -3.80\\
$\#$2 & 14.35 & -7.12 & -6.49 & -6.22 & -6.27 & -6.03 & -5.77 & -5.68 & -5.65\\
$\#$3 & 14.55 & -2.45 & -4.66 & -6.14 & -7.45 & -8.15 & -9.07 & -9.78 & -10.2\\
\hline
\end{tabular}
\end{center}
\end{table*}

In this section we discuss binary progenitor models for type IIb supernova.  We compute the evolution of binary models with different initial orbital periods and different initial mass ratios. We adopt an initial primary mass of 15~\Msun~in agreement with the progenitor model proposed for 1993J \citep{Maund04}. We assume that type IIb supernovae result from massive stars that undergo core collapse with an envelope which contains between 0.1 and 0.5~\Msun~of hydrogen. This criterion is based on observations, previous models \citep{Podslad93,Woosley94, Elmhamdi06} and some test models which proved that a hydrogen mass less than 0.1~\Msun~gives rise to a compact rather than an extended progenitor. 

As an example we discuss a system with an initial orbital period of 1500 days and initial masses of 15 and 14.35~\Msun~for the primary and secondary star respectively.  The initially most massive star evolves faster and experiences significant mass loss in the form of a stellar wind when it ascends the giant branch and during central helium burning.  After about 13.1~Myr, when the helium mass fraction has dropped below 0.5 in the center, it fills it Roche lobe \cite[late case B mass transfer, ][]{Kippenhahn67}.  At this moment it has already lost more than 1~\Msun~and has become less massive than its companion.  The reversal of the mass ratio before the onset of Roche-lobe overflow helps to stabilize the mass transfer.  

In Figure~\ref{MtrMass} we depict the mass-transfer rate as a function of the remaining envelope mass.  We find that the  mass transfer initially takes place on a timescale equal to the thermal timescale of the primary star. The maximum mass-transfer rate during this phase is $6\times10^{-4}~\Msun$yr$^{-1}$. Although this phase lasts only about 0.05~Myr, about 4.5~\Msun~is transferred.  
After this phase the star keeps filling its Roche lobe and mass transfer continues on the nuclear timescale, at a rate comparable to the mass-loss rate in the form of a stellar wind, about $3\times10^{-6}~\Msun$yr$^{-1}$.  During this phase the primary expands on its nuclear timescale while it is burning helium in its center. This phase lasts about 0.8 Myr and about 2.5~\Msun~is transferred.  After central helium exhaustion the star expands again on its thermal timescale and a second maximum in the mass-transfer rate occurs.  Finally, at the onset of carbon burning the star expands again resulting in a third peak in the mass-transfer rate, see Fig.~\ref{MtrMass}.  We follow the evolution of the system up to this point, when the mass of hydrogen in the envelope has decreased to 0.46~\Msun.
Extrapolating the mass-loss rate we find that the amount of hydrogen in the envelope at the time of explosion will be about 0.37~\Msun. Therefore we expect that the primary star explodes as a type IIb supernova. 

In wider systems mass transfer starts in a later phase of the evolution of the primary star.  Because the primary stars in these systems are more evolved, there is less time available to reduce the mass of the envelope before the explosion.    In addition, stellar winds had more time to reduce the mass of the primary star before the onset of Roche-lobe overflow.  Reversal of the mass ratio stabilizes the process of mass transfer. This results in a lower mass-transfer rate during the first phase of mass transfer. Vice versa we find that stars in binary systems with lower initial orbital periods remain with smaller envelope masses at the time of explosion. 
We give details of all our computed models in Tables~\ref{TC}--\ref{TNC} in the Appendix.

\begin{figure*}
\centering
\includegraphics[width=\textwidth]{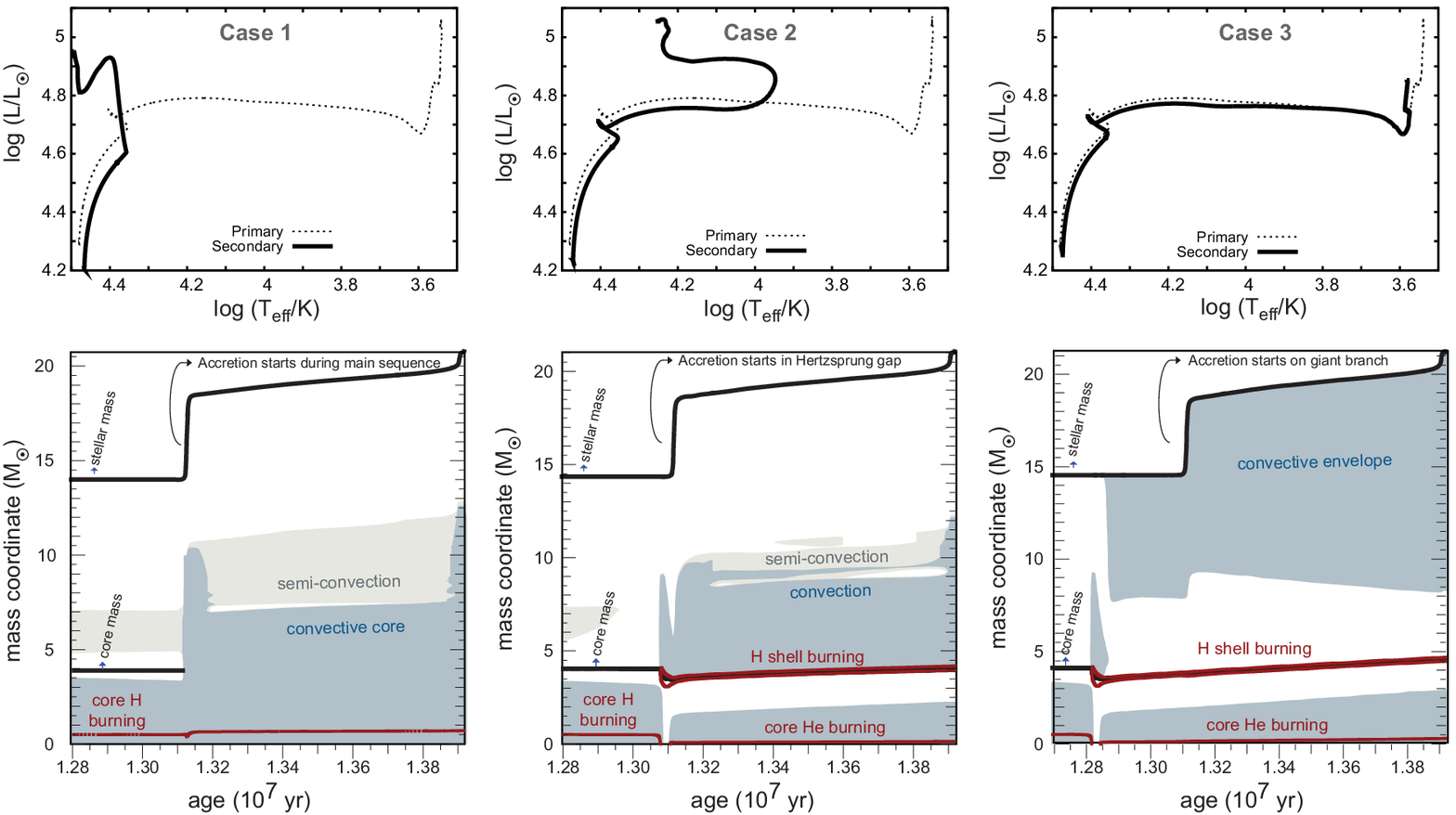}
\caption{Response of the accreting star in three different cases (1) left panels: accretion during main sequence, (2) central panels: accretion during Hertzsprung gap and (3) right panels: accretion during giant branch.  In the top row we show the evolutionary tracks of both stars in the Hertzsprung-Russell diagram. In the bottom row we illustrate the evolution of the internal structure of the accreting star as a function of time around the moment of accretion. The total mass (black line), the core mass (black line) and the mass coordinates of the region in which nuclear burning takes place (red line) are plotted versus time. Grey areas indicate convective regions. In all models we assumed a primary mass of 15~\Msun, an initial orbital orbital period of 1500 days and a secondary masses of 14, 14.35 and 14.55~\Msun~to illustrate the three different cases.  These diagrams correspond to models \#1, \#2 and \#3 in Table~\ref{TC1}.}
\label{all}
\end{figure*}

\subsection{Properties of the companion \label{reaction_companion}}
In general the companion star is relatively unevolved, i.e. still on the main sequence, at the onset of mass transfer.  However, when the initial mass ratio is close to one, the evolutionary timescales of the primary and secondary star are comparable and the companion can be more evolved. The response of the companion to mass accretion depends on its evolutionary stage. 
 We distinguish three different cases:  (1) accretion starts while the companion is on the main sequence, (2) accretion starts while the companion is crossing the Hertzsprung gap and (3) accretion starts while the companion is a giant. 
The evolution of the stars in these three cases is illustrated in Figure~\ref{all}, where we give the evolutionary tracks of both stars in the Hertzsprung-Russell diagram and a Kippenhahn diagram illustrating the changes in the internal structure as a result of mass transfer. Table~\ref{TC1} list several properties of the binary models described here.

The first case, accretion during the main sequence, is depicted in the left panels of Figure~\ref{all}.  The accreting star responds to the increase in mass by adapting its internal structure. The size of its convective core increases and fresh hydrogen is mixed towards the center, effectively rejuvenating the star. After accretion the properties of the star are similar to the properties of a younger single star of the same mass. The star becomes brighter but remains hot, appearing as an O star, see also Table~\ref{TC1} (model \#1).

The central panels of Figure~\ref{all} show an example of the second case, accretion during the Hertzsprung gap (model \#2 in Table~\ref{TC1}). In this phase nuclear burning takes place in a shell around the core. This prevents the star from adapting its internal structure to that of a single star of its new increased total mass.  Having a core mass which is too small compared to the core mass of a normal single star, the star appears as an over-luminous B supergiant.  This type of progenitor model has been proposed to explain the properties of the blue companion of SN 1993J.

For systems with very similar initial masses, accretion takes place while the secondary resides on the giant branch, see the right panels of Fig~\ref{all}. In this third case we find that the secondary will appear as a K supergiant at the moment the primary explodes (model \#3 in Table~\ref{TC1}).  If the initial mass ratio is even closer to one we find that the evolution of the secondary is accelerated enough to catch up with and overtake the evolution of the primary. In this case the secondary explodes first, as a normal type II supernova, while the primary explodes afterwards as a type IIb supernova. The time difference between the explosions is 8000 years for model \#4 in Table~\ref{TC1}, but can be up to $10^5$ years for very close mass ratios (see Table~\ref{GTC} in the Appendix). 

While the three companions of the SN IIb progenitors in these examples are of similar luminosity, it is important to consider how easily they might be observed in pre- and post-explosion images. For example the O star has a high surface temperature and so most of its emission is in the ultraviolet. In Table~\ref{TUV} we list the expected broad-band magnitudes of our example progenitors using the methods outlined in \cite{Eldridge2007} and \cite{Eldridge2009} to calculate the colours. We see that while all three progenitors are of a similar bolometric luminosity, the B and K supergiants output more of their light in the optical bands, such as V and I which are typical of those most commonly available in pre-explosion imaging \citep[]{Smartt09}, making them easier to identify in pre- and post-explosion images.

\subsection{Parameter space}
\label{sec:parameterspace}

\begin{figure*}
\begin{center}
\includegraphics[width=12cm]{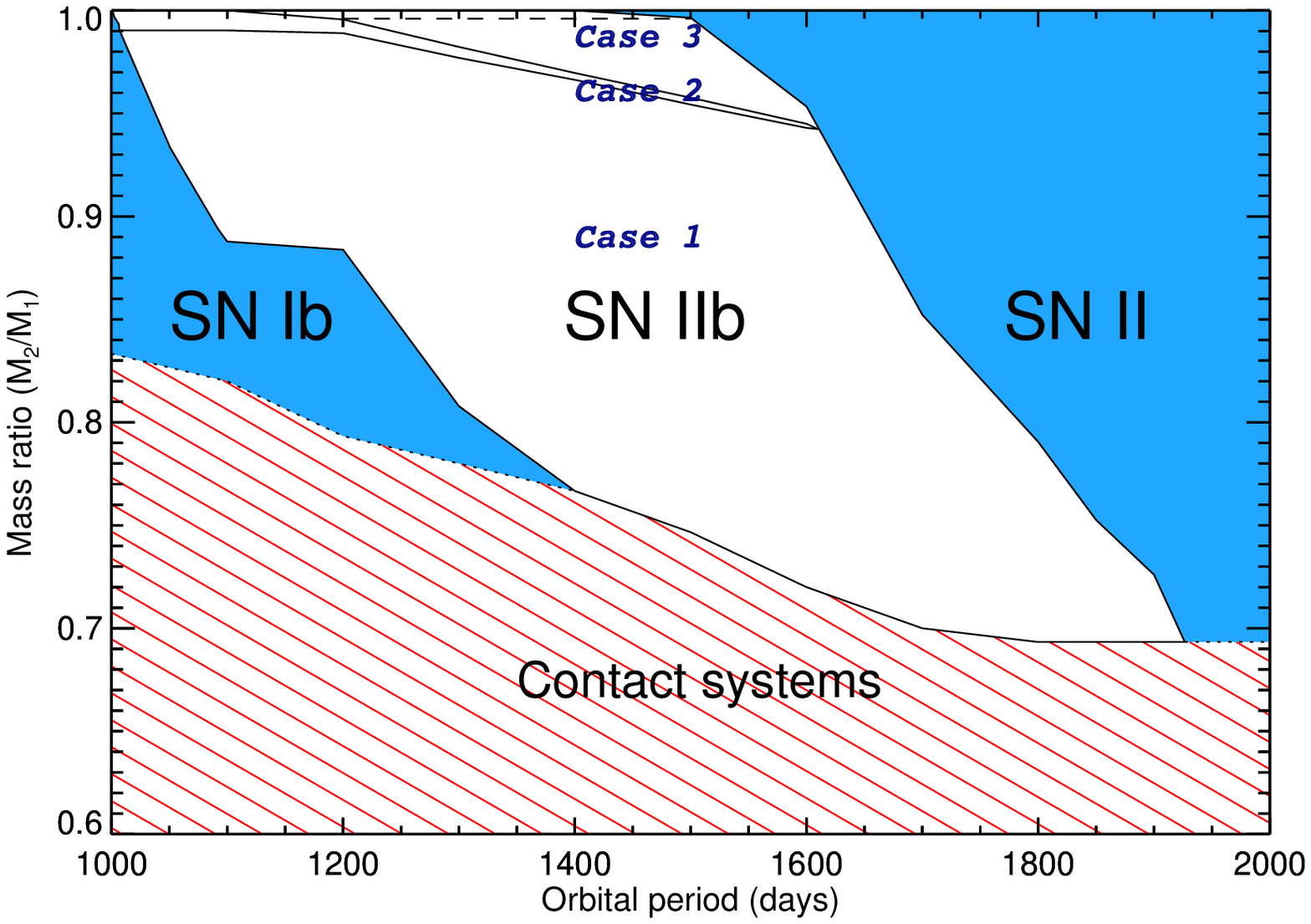}
\caption{Ranges of initial mass ratios and orbital periods of binaries for which the primary star is expected to explode as a type IIb of the extende type supernova.  We assumed a primary mass $M_1 = 15$~\Msun, a metallicity $Z=0.02$ and conservative mass transfer. 
We assume that type Ib supernovae result from stars with less than 0.1~\Msun~of hydrogen in their envelope at the time of explosion, type IIb supernovae from stars with 0.1--0.5~\Msun~of hydrogen and type II with more than 0.5~\Msun~of hydrogen.  Furthermore we distinguish different cases based on the response of the companion star. Case 1: The secondary evolves to an O star at time of explosion of the primary. Case 2: the secondary evolves to a B supergiant. Case 3: the secondary evolves to a K supergiant. See also Fig.~\ref{AreaZoom} for an enlargement.  }
\label{Area}
\end{center}
\end{figure*}

Figure~\ref{Area} depicts the range of initial mass ratios and initial orbital periods of binary systems in which the primary star is expected to explode as a type IIb supernova.   
In systems with initial orbital periods larger than about 1600--1800 days, depending on the mass ratio, the primary star has more than 0.5~\Msun~of hydrogen left in its envelope at the time of explosion. We assume that the supernova would be classified as type II. We therefore do not find any SNe type IIb progenitors which have undergone Case C mass transfer \cite[]{Lauterborn70}, since these all end up with hydrogen masses greater than 0.5~\Msun. In systems with initial orbital periods smaller than about 1000--1300 days the envelope mass left at the time of explosion is less than 0.1~\Msun~and we assume that the primary explodes as a type Ib supernova. 
In systems with $M_2/M_1 \lesssim$ 0.7--0.8, depending on the orbital period,  the mass-transfer rate is so high that the stars come into contact.  These systems are expected to experience a common envelope phase. As we have discussed in section~\ref{sec:model}, whether or not a binary evolves into contact during this phase is sensitive to some of our model assumptions. Therefore the critical mass ratio separating contact from non-contact systems is uncertain and we regard the location of this boundary in Fig.~\ref{Area} as indicative only.

The borders between type IIb / II and between type Ib / IIb run diagonally across this diagram, i.e. the critical orbital period dividing different supernova types increases with mass ratio.  This is caused by the fact that systems with more extreme initial mass ratios exhibit a higher mass-transfer rate in the initial phase of mass transfer. For a given initial orbital period this results in lower envelope masses at the time of explosion in systems with more extreme mass ratios.  

The border between type Ib and type IIb shows a horizontal step near mass ratios of about 0.89.  This feature is related to the evolutionary stage of the primary star at the onset of mass transfer.  
For systems with orbital periods smaller than about 1100--1200 days,  the primary star fills its Roche lobe relatively early, while it is still in the Hertzsprung gap just before ascending the giant branch.  The effects of stellar winds, which enlarge or even reverse the mass ratio and stabilize the mass transfer, are still limited.  In addition, the rapid expansion of the primary on its thermal timescale results in a high mass-transfer rate. We find that within this period range only in systems with very similar initial masses, $M_2/M_1 \gtrsim 0.89$, enough hydrogen can be retained on the surface of the primary star until the time of explosion to result in a type IIb supernova.  

\begin{figure*}
\begin{center}
\includegraphics[width=12cm]{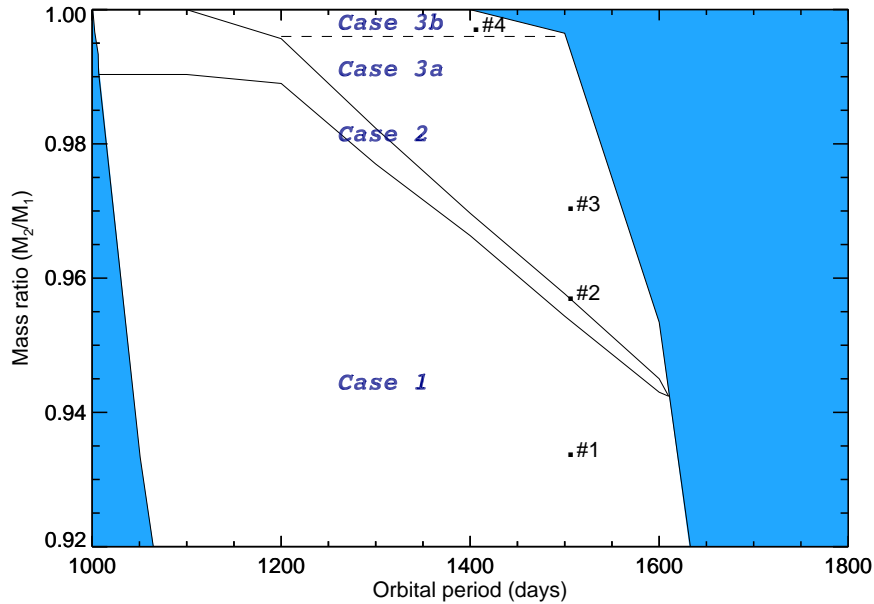}
\caption{Zoom-in of Fig. \ref{Area} for initial mass ratios close to one. The region (case~3) where the secondary evolves to a K supergiant is subdivided into case 3a, where the primary explodes before the secondary, and case 3b, where the secondary explodes before the primary. The location of the four models in Table \ref{TC1}) is also indicated.}
\label{AreaZoom}
\end{center}
\end{figure*}

In Figure~\ref{Area} we also indicate the three different cases distinguishing  the properties of the secondary star at the moment of explosion, discussed in Sect.~\ref{reaction_companion}.
The information from our model grid on which this figure is based can be found in Tables~\ref{TC}--\ref{GTC} in the Appendix.
In the most common case~1, the companion is still on the main sequence at the onset of accretion and it will appear as an O star at the time of explosion.  Case~2 is the region where the companion accretes while it is on the Hertzsprung gap and will appear as a B supergiant.  The range of mass ratios for this case is very limited. This is a direct result of the short time spent by the star in the Hertzsprung gap, about 0.5\% of time that it spends on the main sequence.   Case~3 indicates the region where the companion will evolve to a K supergiant.  This case is more likely to occur for wider systems in which the primary star fills its Roche lobe in a later stage of helium burning.  In Figure~\ref{AreaZoom} we show an enlarged region of the parameter space for initial mass ratios near 1.  Here we indicate case~3b in which the secondary explodes before the primary. This only occurs if the stars are initially very close in mass, to within 0.5\%.

\subsection{Non-conservative mass transfer}

\begin{table*}
\caption{Properties of binary stellar models for an initial primary mass of 15~\Msun, with an initial orbital period of 1500 days and with variation of the initial secondary mass M$_{\rm 2i}$ and the accretion efficiency of the companion $\beta$.} 
\label{TNC1}
\begin{center}
\begin{tabular}{|c|c|c|c|c|c|c|c|c|c|c|c|c|}
\hline \hline 
$N^0$ & 
\begin{tabular}{c}
$M_{2i}$\\($M_\odot$)\\ 
\end{tabular} 
& \begin{tabular}{c}
$M_{1f}$\\($M_\odot$)\\ 
\end{tabular} 
& \begin{tabular}{c}
$M_{2f}$\\($M_\odot$)\\ 
\end{tabular} 
& $\beta$
& \begin{tabular}{c} 
$M_{H1}$\\($M_\odot$)\\
\end{tabular} 
& log$\left(\frac{L_1}{L_\odot}\right)$ & 
\begin{tabular}{c}
$T_{\rm eff,1}$ \\($10^3$K)\\
\end{tabular} 
& log$\left(\frac{L_2}{L_\odot}\right)$ & 
\begin{tabular}{c}
$T_{\rm eff,2}$ \\($10^3$K)\\
\end{tabular}\\
	\hline 
$\#$1 & 14 & 5.93 & 20.92 & 1.00 & 0.354 & 5.07 & 3.48 & 4.95 & 31.4\\
$\#$1a & "" & 6.38 & 17.23 & 0.50 & 0.598 & 5.07 & 3.40 & 4.43 & 27.1\\
$\#$2 & 14.35 & 5.98 & 21.23 & 1.00 & 0.370 & 5.07 & 3.47 & 5.06 & 17.9\\
$\#$2a & "" & 6.06 & 20.46 & 0.90 & 0.416 & 5.07 & 3.45 & 5.04 & 16.6\\
$\#$2b & "" & 6.14 & 19.72 & 0.80 & 0.462 & 5.07 & 3.43 & 5.01 & 14.3\\
$\#$2c & "" & 6.19 & 19.35 & 0.75 & 0.489 & 5.07 & 3.43 & 4.97 & 8.67\\
$\#$2d & "" & 6.24 & 18.97 & 0.70 & 0.518 & 5.07 & 3.42 & 4.84 & 3.77\\
$\#$2e & "" & 6.34 & 18.25 & 0.60 & 0.575 & 5.07 & 3.40 & 4.81 & 3.78\\
$\#$2f & "" & 6.45 & 17.54 & 0.50 & 0.633 & 5.07 & 3.39 & 4.81 & 3.77\\
$\#$2g & "" & 6.74 & 15.87 & 0.25 & 0.806 & 5.07 & 3.37 & 4.80 & 3.72\\
$\#$3 & 14.55 & 6.01 & 21.40 & 1.00 & 0.390 & 5.07 & 3.45 & 4.85 & 3.82\\
$\#$3a & "" & 6.49 & 17.72 & 0.50 & 0.660 & 5.07 & 3.39 & 4.85 & 3.79\\
$\#$A& 8.6 & 5.41 & 10.44 & 0.25 & 0.091 & 5.07 & 3.79 & 3.93 & 25.4\\
$\#$B & 8.8 & 5.41 & 12.49 & 0.50 & 0.096 & 5.07 & 3.80 & 4.17 & 27.8\\
$\#$C & 9.7 & CONTACT & / & 0.75 & / & / & / & / & /\\
$\#$D & 10 & 5.50 & 15.49 & 0.75 & 0.131 & 5.07 & 3.69 & 4.46 & 30.5\\

\hline
\end{tabular}
\end{center}
\tablefoot{The final mass of the primary and secondary, M$_{\rm f1}$ and M$_{\rm f2}$, and the amount of hydrogen in the envelope M$_{\rm H}$ are given at the time of explosion, together with the luminosity log(L/L$_\odot$) and effective temperature T$_{\rm eff}$ of the primary and secondary at the onset of central carbon burning. If the evolution of the binary system ended because of formation of a contact binary, the parameters cannot be determined and are indicated by '/'.}
\end{table*}

\begin{figure}
\includegraphics[width=6cm,angle=270]{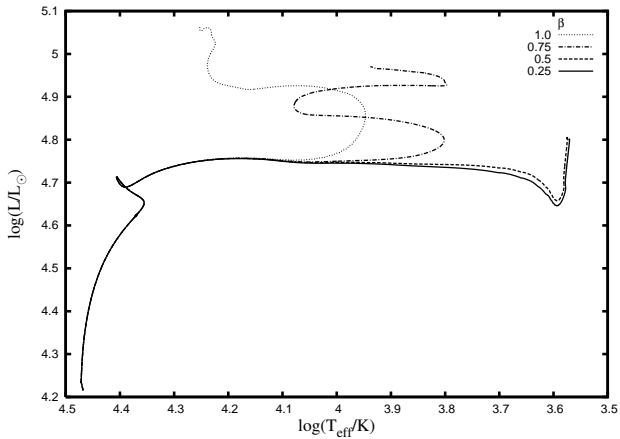}
\caption{Evolution tracks in the Hertzsprung-Russel diagram of the secondary of binary systems with initial mass 15+14.35 M$_\odot$ and initial orbital period 1500 days. The differences between the evolution tracks of the secondary are caused by variation of the accretion efficiency ($\beta$). (see models $\#$2 - $\#$2g in Table \ref{TNC1})}
\label{CBRHR}
\end{figure}

So far we have assumed that the mass transferred during Roche-lobe overflow is efficiently accreted by the secondary.  This assumption may not be valid, given the high mass-transfer rates reached and the fact that the secondary star may quickly spin up to break-up rotation after accreting only a fraction of the transferred mass \cite[e.g.][]{Packet81}. In this section we investigate the effect of variations in the accretion efficiency $\beta$, i.e. the fraction of material accreted by the secondary with respect to the amount of material lost by the primary star as a result of Roche-lobe overflow. 

The effect on the primary star is rather modest as its evolution has been determined largely by its development before the onset of Roche-lobe overflow.  In Table~\ref{TNC1} we list various properties of models computed with accretion efficiencies varying  between 1 and 0.25.  The luminosity and temperature of the primary star at the moment of explosion are hardly affected; in all cases we find that the primary star will be a cool supergiant at the time of explosion. Nevertheless its final mass and therefore the amount of hydrogen in the envelope at the moment of explosion $M_{\rm H1}$ increases if we assume a lower accretion efficiency. This can be understood as an effect related to response of the orbit and thus the size of Roche lobe to mass transfer and mass and angular-momentum loss from the system.  In our models mass loss from the system widens the orbit and therefore the Roche lobe. This results in smaller mass-transfer rates and therefore larger envelope masses at the time of explosion.   This influences the parameter space for which we predict type IIb supernovae, as depicted in Fig.~\ref{Area}.  For an accretion efficiency of 50\% the range of initial orbital periods resulting in SN IIb progenitors shifts by about 200 days to smaller periods in comparison with the conservative case.

The evolution and properties of the secondary are more sensitive to the adopted accretion efficiency. The lower accretion rate reduces the expansion of the companion star.  As a result, the formation of contact can be avoided in systems with more extreme initial mass ratios. This effect widens the parameter space for type IIb supernovae.  In table~\ref{TNC1} we list a few test models that explore the effect of nonconservative mass transfer for extreme mass ratios.  For example, we find that the mass ratio leading to contact for a system with an initial orbital period of 1500 days shifts from $M_2/M_1 \approx 0.74$ assuming conservative mass transfer to about 0.65  if we assume a accretion efficiency of 75\% (see models $\#$C-$\#$D in Table~\ref{TNC1}). 
For lower accretion efficiencies we find that formation of contact is no longer the effect that limits the parameter space for type IIb progenitors.  In systems with such extreme mass ratios, that mass stripping from the primary star becomes so efficient that these stars do not have enough hydrogen left at the moment of explosion to become IIb. The most extreme mass ratio resulting in a IIb progenitor assuming the same initial orbital period shifts to 0.59 (0.58) assuming an accretion efficiency of 50\% (25\%), see model $\#$B, ($\#$A) in Table~\ref{TNC1}.  In section~\ref{sec:rates} we discuss the influence of $\beta$ on the rate of type IIb supernova.

The appearance of the companion star at the moment of explosion is also affected.  In the first case of accretion during the main sequence the secondary rejuvenates, regardless of the adopted accretion efficiency. However, the effective temperature and luminosity decrease for lower accretion efficiencies as a result of the smaller amount of accreted mass (e.g. compare models \#1 and \#1a in Table~\ref{TNC1}). 
The evolution of the secondary in the second case of accretion during the Hertzsprung gap depends most strongly on the accretion efficiency. In Fig. \ref{CBRHR} we illustrate the evolutionary tracks of the secondary in the Hertsprung-Russel diagram.  For conservative mass transfer the secondary evolves to an overluminous B star. The effective temperature decreases with decreasing accretion efficiency. If the accretion efficiency is 75\%, the secondary evolves to an A supergiant, and for accretion efficiencies of 50\% or less the secondary evolves to a much cooler K supergiant (e.g. compare models \#2 and \#2g in Table~\ref{TNC1}).
In the third case of mass transfer while the secondary is already on the giant branch, the secondary still evolves to a K supergiant. In this situation its luminosity and effective temperature are almost the same as for conservative mass transfer (e.g. compare models \#3 and \#3a in Table~\ref{TNC1}).
 
In summary we conclude that lower accretion efficiencies shift and widen the parameter space for type IIb supernovae from stable mass transfer. However, it proves to be even more difficult to produce a companion that is a B supergiant at the moment of explosion for non-conservative mass transfer. If the accretion efficiency is smaller than about 60\%  our models predict no companions to reside in the middle of the Hertzsprung-Russel diagram. They will either be hot and compact residing in the main sequence band or, if the initial masses were very similar, the secondary will appear as a K supergiant.

\begin{table*} 
\caption{Predicted rates for type IIb supernovae from stable mass transfer in binaries and its dependence on different input assumptions.}
\label{tab:rates}
\begin{center}
\begin{tabular}{rccc}
\hline \hline   
Mass ratio distr.  & { fav. extreme ratios} & {\bf flat}  & fav. equal masses \\

 & \multicolumn{1}{c}{$(x =-1)$} & \multicolumn{1}{c}{$\mathbf{(x=0)}$}  & \multicolumn{1}{c}{$(x = 1)$} \\
\hline
\multicolumn{4}{l}{ {Fraction of binaries producing type IIb SNe via stable mass transfer}}\\
  & \multicolumn{1}{c}{0.83\%} & \multicolumn{1}{c}{1.33\%} & \multicolumn{1}{c}{1.85\%}  \\

&&&\\
\multicolumn{4}{l}{ {Number of type IIb relative to the number of all core collapse SNe}}\\
\multicolumn{4}{l}{ {(assuming that the companion does -- does not produce a core collapse SNe)}}\\
  
$f_{\rm bin} = 25\%$  & 
0.17 -- 0.21\% &  
0.27 -- 0.33\% & 
0.37 -- 0.46\% \\
{$\mathbf{f_{\rm bin} =50\%}$}  & 
0.28 -- 0.42\% & 
{\bf 0.44 -- 0.67\%} & 
0.62 -- 0.93\% \\
$f_{\rm bin} =100\%$  & 
0.42 -- 0.83\% & 
0.67 -- 1.33\% &  
0.93 -- 1.85\% \\
&&&\\
 \multicolumn{4}{l}{ {Estimates of the rates assuming lower accretion efficiencies $\beta$}$^1$}\\

$\mathbf{\beta=1.0}$ & & {\bf 0.44 -- 0.67\%} &  \\
$\beta=0.75$ & & 0.60 -- 0.92 \%$^{2}$ & \\
$\beta=0.5$ & & 0.69 -- 1.06\%$^{2}$ & \\
$\beta=0.25$ & & 0.71 -- 1.09\%$^{2}$ & \\
&&&\\

\multicolumn{4}{l}{{Assuming that all systems with orbital periods between 1000 and 2000 days produce IIb's$^{1,3}$}}\\
 & - & 2.3--3.5\% & - \\
 \hline
\end{tabular}
\end{center}
$^1$Assuming a binary fraction of $f_{\rm bin} =50\%$. \\
$^2$The estimates for lower accretion efficiencies are based on just a few test models and should only be taken as indicative.   A more extended grid of non-conservative models is required for a proper evaluation of these rates. \\ 
$^3$This estimate does no longer depend on the assumed distribution of initial mass ratios. 
\tablefoot{See Section~\ref{sec:rates} for details. The rate predicted based on our standard assumptions is marked in boldface.}
\end{table*}

\begin{table*} 
\caption{Relative rates for the  properties of the companion star at the moment of explosion for different initial mass ratio distributions based on conservative models.}
\label{tab:rates2}
\begin{center}
\begin{tabular}{rccc}
\hline \hline   
Mass ratio distr.  & { fav. extreme ratios} & { flat}  & fav. equal masses \\

 & \multicolumn{1}{c}{$(x =-1)$} & \multicolumn{1}{c}{$\mathbf{(x=0)}$}  & \multicolumn{1}{c}{$(x = 1)$} \\
\hline
 
Case 1 (O star) & \multicolumn{1}{c}{91.5\%} & \multicolumn{1}{c}{90.3\%} & \multicolumn{1}{c}{89.1\%}  \\
Case 2 (B supergiant) & \multicolumn{1}{c}{2.5\%} & \multicolumn{1}{c}{2.8\%} & \multicolumn{1}{c}{3.2\%}  \\
Case 3 (K supergiant) & \multicolumn{1}{c}{6.0\%} & \multicolumn{1}{c}{6.9\%} & \multicolumn{1}{c}{7.7\%}  \\

\hline
\end{tabular}
\end{center}

\end{table*}

\section{Predicted rates \label{sec:rates}}

The currently most likely observed rate of type IIb supernovae with respect to all core collapse supernova is about 3 \% (see Sect.~\ref{sec:obs} for a discussion).
The accuracy of the observed rate is expected to go up in the near future thanks to the automated detection and classification of supernova light curves.  Computing the rate predicted from our models is not straightforward and requires adopting further uncertain assumptions about for example the binary fraction and the distribution functions of the initial parameters.  That said, we still consider it worthwhile to estimate the rate predicted from our models, which we do below adopting commonly made, most reasonable assumptions and assessing the effect of uncertainties in these assumptions on the rates. 

In the previous sections we discussed progenitor models assuming an initial mass of  15~\Msun~for the primary star.  Based on a few test models we conclude that the parameter space does not significantly change for systems with primary masses 
between roughly 10 and 20 solar masses. This mass range dominates the population of massive binaries in which the primary is massive enough to undergo a core collapse supernova.  In the following derivations we will assume that the characteristics of our conservative models with a 15~\Msun~primary are representative for the population of massive binaries. 

We assume that the initial orbital periods, $P$, are distributed according to \"Opik's \citeyearpar{Opik1924} law \cite[]{Kouw07},
\[
f(P)\propto P^{-1}\quad \quad {\rm for} \quad \quad  0.5 \le P\,(\rm{d}) \le 10^4. 
\]
The upper limit is chosen close to the widest orbital period for which we expect binaries to interact via Roche-lobe overflow during their life time.  For the initial mass ratio, which we define as the mass of the initially less massive star over the mass of the initially most massive star,  $q\equiv M_2/M_1$, we adopt a power-law distribution, 
\[
f\left( q  \right)   \propto q^{x} \quad \quad {\rm for} \quad \quad 0.25 \le  q   \le 1.  
\]
  Binaries with extreme mass ratio are hard to detect and even if they have been detected it is still difficult to determine their mass ratio accurately. Therefore we limit ourselves to systems with  $M_2/M_1 > 0.25$, following the approach of for example \citet{Pols1991}.    
The parameter  $x$ describes whether the distribution of initial mass ratios is flat ($x=0$), skewed towards systems equal masses ($x>0$) or favors systems with unequal masses ($x<0$).   
Although these distribution functions are uncertain, they are consistent with observed distributions. \cite{Kouw07} and references therein find $x= -0.4$ for the nearby association Scorpius OB2 while  $x=-1$ corresponds to a study of spectroscopic binaries by \cite{Trimble90}.  A flat or uniform mass ratio distribution $x=0$ has been  quoted by \cite{Sana+09} for young open clusters, whereas e.g.\ \citet{Pinsonneault+Stanek06} and  \citet{Kobulnicky+Fryer07} claim that massive binaries like to be twins, i.e.\ $x>0$.  We will adopt $x=0$ as our standard assumption and consider the two extreme cases $x=\pm1$ as well.

We adopt a binary fraction $f_{\rm bin}$ of 50\% as our standard assumption, i.e. for every single star with mass $M_1$ there is exactly one binary system of which the primary mass is $M_1$ and the orbital period and mass ratio are within the ranges specified above.  For comparison,  \citet{Mason+98} derives a binary fraction of 59\%-75\% for O stars in clusters and associations.  \citet{Garcia+Mermilliod01} find fractions between 14-80\%, whereas  \cite{Kobulnicky+Fryer07} infer a fraction higher than 70\%. On the other hand, Pols et al (1991) adopted a fraction of 27.5\% of binaries with B-type primaries, $P < 10$\,yr and $0.25 < q < 1$. Besides our standard assumption we will consider binary fractions of $f_{\rm bin} = 25\%$ and 100\%. We note that exploring the effect of different binary fractions partially covers the uncertainties in the assumed upper limits for the orbital period and mass ratio that we assumed above  for the initial parameter distributions. 

The observed rate of type IIb supernovae is expressed with respect to the rate of all types of core collapse supernova. To compute this number we must consider that a significant fraction of the companion stars end their lives as core collapse supernovae.  The precise fraction will depend on the initial mass ratio distribution, the amount of mass lost from the secondary and the fraction that is actually  accreted by the secondary.  For clarity, we will just consider the two extreme cases in which all or none of the companion stars result in a core collapse supernova.

\subsection {The rate of type IIb supernovae}

Under the standard assumptions described above, we find that the number of type IIb supernovae over the number of core collapse supernovae formed via stable mass transfer as predicted from our models is roughly 0.6 \%, see Table~\ref{tab:rates}, about a factor five lower than the observed rate (Sect.~\ref{sec:obs}). The rate is larger, but only by a factor two, than the rate predicted from single stars, 0.3\%, under the assumption that all stars are single (see Section~\ref{sec:single}).  
Our predicited rate is compatible with the rate found by \cite{Podsiadlowski92} for the 'stripped' supernovae. In contrast to what some previous authors have suggested \citep[e.g.][]{Podslad93}, we emphasize that not only for the single star channel but also for the binary channel one needs to fine-tune the initial parameters of the system. For single stars the initial mass must be finely-tuned, whereas for binaries the combination of initial orbital period and inital mass ratio must by be carefully arranged.

The derived rate depends on the adopted initial mass ratio distribution, the assumed binary fraction and on whether or not the companion star results in a core-collapse supernovae, see Table~\ref{tab:rates}. Even when we adopt the assumptions favoring a high SN IIb rate from binaries (100\% binaries, mass ratio distribution skewed to equal-mass systems and no core-collapse supernovae from the companions) the rate derived from our conservative models is still below the observed rate.

Another uncertainty to consider is the efficiency of mass transfer. In our models, lower efficiencies shift and widen the parameter space. In Table~\ref{tab:rates} we show that the number of SNe IIb over the number of core-collapse supernovae may increase to over 1\%, with the warning that for a proper evaluation of the parameter space one needs a more extensive model grid. 

A further uncertainty we have to consider is in the range of hydrogen-envelope masses in the progenitor that give rise to a SN IIb. 
The lower limit of 0.1~\Msun~is determined by the boundary between compact and extended type IIb. The upper limit of about 0.5~\Msun~is determined by the lack of a plateau in the light curve, which depends on the chemical structure of the stellar envelope and therefore is less certain. If we change the boundary from 0.5 to 0.6~\Msun, test models show that the initial orbital period range will only widen by about 80 days. This means an increase of the overal rate of type IIb SNe of about 10\%. The uncertainties in these boundaries will therefore not have a large effect on the rate.

As an extreme assumption, we consider that all systems with orbital periods between 1000 and 2000 days produce IIb progenitors.  This corresponds to 7\% of all binary systems with initial mass ratios and orbital periods in the ranges specified earlier.  This would increase the relative rate compared to the rate of core collapse supernova to 2.3--3.5\%, depending on whether the companion star produces a core-collapse supernova or not and assuming a binary fraction of 50\%.  This number is consistent with the observed rate.

\subsection {The relative rates of different cases for the companion \label{{sec:rates:comp}}}
Table~\ref{tab:rates2} indicates the relative rates for the different characteristics of the companion at the moment of explosion, as described in section~\ref{reaction_companion}.  The large majority, about 90\%, of type IIb supernova resulting from stable mass transfer is expected to have an O-star companion at the moment of explosion.  The rate of systems with a blue supergiant companion, which applies to SN 1993J and possibly 2001ig, is predicted to be quite low according to our models, about 3\% of the type IIb supernova rate. These relative rates reflect the large and very small regions in parameter space for O-star and B-supergiant companions, respectively, as depicted in Fig.~\ref{Area}. However, these percentages do not necessarily reflect the probability of \emph{detecting} a companion of a certain type. Detecting an O-star companion in post-explosion images of a supernova is more difficult than detecting a B or K supergiant, as we demonstrated in Section~\ref{reaction_companion}.

The probability for the presence of blue companions to type IIb SNe we derive is strikingly lower than predicted by \citet{Podsiadlowski92}.  They find that, if the companion starts to accrete after becoming a giant star, it will contract in a similar way to a feature known as ``blue loops'' seen in some evolutionary tracks of single stars. We do not find this behavior with our stellar evolution code. Whether or not stellar models perform blue loops is very sensitive to the ratio of core mass to total stellar mass and to details in the chemical profile outside the stellar core \cite[]{Kippenhahn94}. If a significant fraction of the case 3 models would result in blue companions, the relative rate for the presence of blue companions may increase by a factor of two or three (see Table~\ref{tab:rates2}). 

In the most common case, accretion during the main sequence, the star rejuvenates, i.e. the size of its convective core increases to adapt to the new stellar mass, see section~\ref{reaction_companion}. In our code, all main sequence stars that accrete rejuvenate.  However, whether stars rejuvenate or not depends on the assumed efficiency of semi-convection.  Stars that have evolved towards the end of the main sequence build up a chemical gradient between the helium rich core and the hydrogen rich envelope. \citet{Braun95} show that, assuming a low efficiency of semi-convective mixing, the growth of the convective core in an accreting main-sequence star is prevented. These stars may also appear as blue supergiants at the moment of explosion.  

Based on their models, we estimate that in our case of conservative mass transfer the companion must have reached a central helium mass fraction of about 0.75 or more to prevent rejuvenation. In our models this occurs for initial mass ratios larger than about 0.87, and we estimate that this may increase the relative rate of the presence of blue companions up to about 50\%.    For lower accretion efficiencies it becomes even easier to prevent rejuvenation: a central helium mass fraction for the accreting star of just 0.6 and higher are required if $\beta=0$.  However, the companion must accrete enough to finish its main sequence evolution before the explosion of the primary star, to be observed as a blue supergiant.  More detailed estimates are beyond the scope of this study, but we do  emphasize this as an interesting possibility to use the companions of supernovae to gain insight in internal mixing processes.

\section{Discussion \& Conclusion \label{sec:disc}}

We identified binary progenitor models for extended type IIb supernovae. In these models the most massive star is stripped from its envelope by interaction with its companion, such that only several tenths of solar masses of hydrogen remain at the moment of explosion.  We find that the most massive star must fill its Roche lobe during central helium burning in order to achieve this. We derive for which range of initial orbital periods and initial mass ratios binary systems are expected to produce type IIb supernova progenitors. 

We have discussed in detail the properties of the companion star at the moment of explosion, motivated by the detection of a companion for SN 1993J, a possible companion for 2001ig and the non-detection of a companion in the supernova remnant of Cas A.  
We distinguish three cases: (1) the companion appears as a hot O star, when it accreted while still being on the main sequence, (2) the companion becomes an over-luminous B-star if mass transfer started while the companion was crossing the Hertzsprung gap, a scenario which may apply to 1993J and possibly 2001ig and (3) the companion will be a K-supergiant when it accreted after ascending the giant branch. 
The third case applies to systems with very similar initial masses and can be subdivided in (3a) systems in which the initially most massive star explodes first and (3b) systems in which the companion star explodes first, up to $10^5$ years before the primary. These models may apply to Cassiopeia A -if it was an extended-IIb SN- explaining the lack of evidence for a companion\footnote{Signatures from the first supernova, for example in radio emission or from interaction with the interstellar medium will be hard to detect as Cas A is bright radio source itself and the remnant of the first supernova will be roughly 10 to 100 times  larger than the remnant of Cas A.}. If the progenitor of Cas A was compact, not a red supergiant, at time of explosion another evolutionary scenario then discussed here could explain this specific supernova remnant, e.g. a common envelope scenario.

However, our models predict that the scenarios we propose for 1993J, 2000ig (case 2) and Cas A (case 3b) are very rare and require that the companion efficiently accretes the transferred material.  The accretion efficiency, i.e. the fraction of transferred material that is accreted by the companion,  is a major uncertainty for binary evolutionary models. A lower accretion efficiency shifts the parameter space to smaller orbital periods (by about 200 days for an efficiency of 50\% compared to conservative mass transfer) and widens the parameter space towards  more extreme initial mass ratios (assuming an efficiency of 50\% we estimate that the IIb rate predicted by our models increases by roughly a factor 1.6 compared to conservative mass transfer). However a B supergiant is identified more easily in comparison with the other possible evolutionary paths of the companion star, as pointed out in section~\ref{reaction_companion} and Table~\ref{TUV}, which gives a bias towards a higher fraction of observed B-supergiants as a companion star.

Our conservative models predict  $\sim0.6\%$ of all core collapse supernova to be of type IIb (using our standard assumptions, see Section ~\ref{sec:rates}), about a factor of five lower than the currently most likely observed rate.  This rate is larger, but only by a factor of two than the rate predicted for a pure single star population, $\sim0.3\%$, in which high mass stars lose their envelope due to stellar winds.  We emphasize that both the binary and single stars scenario only produce type IIb in a very limited region of the initial parameter space. 

We warn the reader that the rates quoted above are uncertain. The rate from single stars is affected by our limited understanding of mass loss rates from massive red giants. The rate from binary stars is affected by uncertainties in the physics of binary interaction as well as the adopted distributions of initial binary parameters.  If we make several assumptions that favor a high rate of IIb supernovae (a high binary fraction, a distribution of initial mass ratios which is skewed towards systems with equal masses, a low mass-transfer efficiency and only moderate angular momentum loss from the system), we derive a rate that is consistent with the lower limit  of the currently most reliable observed rate of \cite{Smartt09}. 

In addition, the observed rate is based on only a few SNe and it is difficult to make a distinction between an extended type IIb and a compact, the latter of which is more similar to a type Ib SN and is formed by another evolutionary scenario \cite[e.g. ][]{YoonWL10} . Therefore, the observed rate is also one of the uncertainties. Nevertheless, the rates predicted from our standard models seem to indicate that there is definitely room for a single star scenario \emph{and} a common envelope scenario leading to a type IIb.

\subsection*{ A brief comparison with previous work}

Our findings are consistent with earlier studies by \citet{Podsiadlowski92, Podslad93}, \citet{Woosley94}, \citet{Maund04} and \citet{Stancliffe09}.  Nevertheless we can note two major differences.  Contrary to suggestions by \citet{Podslad93} we find that it proves to be hard to explain the presence of a blue companion star at the time of explosion, confirming the thought raised  by \citet{Stancliffe09} with our more extended model grid.   In addition, we find that the primary star needs to fill its Roche lobe during helium burning (late Case B mass transfer), whereas \citet{Podslad93} require mass transfer to start after helium burning (Case C mass transfer).  This can be attributed to differences in the adopted physics: the adopted initial composition, the opacity tables and the mixing length parameter which have a not negligible effect on the radii of giants and therefore on the onset of and response to Roche-lobe overflow.  Furthermore there are small differences in prescription for the mass-transfer rate, the treatment of non-conservative mass transfer and the mass loss occurring between the last computed model and the explosion.  As a result we find that slightly different initial orbital periods are required to produce a IIb progenitor.  However, if we allow for uncertainties, our estimate of the IIb rate is consistent with previous studies.

 \subsection*{What type IIb SNe teach us about stellar and binary physics}

The rate of type IIb supernovae and the relative occurrence of different companions have the potential to constrain both stellar physics, such as internal mixing processes and stellar wind mass loss, but also the physics of interacting binaries, when the accuracy of the determined rates increases in the near future. 

The discrepancy between the currently most likely observed rate and the rate predicted by our models seems to point towards lower accretion efficiencies.  At lower accretion efficiencies, systems with more extreme mass ratios can avoid contact and retain enough hydrogen in their envelope to become progenitors of IIb supernovae. On the other hand, the presence of a blue companion for SN 1993J and possibly SN 2000ig  indicates that at least a substantial fraction must be accreted. Therefore, improving statistics about IIb supernovae may in the future help to constrain the efficiency of mass transfer, one of the major uncertainties in binary evolutionary models.    

The need to avoid contact and a subsequent spiral in for these binaries with an evolved giant companion also gives information about two other uncertain input assumption related to the physics of mass transfer. It may indicate that the entropy of accreted material must be low, to prevent swelling of the companion, and it may indicate that the mass-transfer rate in such systems remains lower than the self-regulated thermal-timescale rate, see section~\ref{sec:model}. 

Finally the detection of one and possibly two blue companions while our standard models predict that this should be a rare event is puzzling.  As discussed in section~\ref{{sec:rates:comp}}, this may be explained with a low efficiency of semi-convection.

 \subsection*{Outlook}
We illustrated the potential of type IIb supernovae,  being on the border between the hydrogen rich type II and the hydrogen poor type Ib/c supernova, to give insight in stellar physics such as internal mixing processes and stellar winds as well as physics of interacting binaries.
Even though the predictions from stellar evolutionary models are still challenged by various uncertainties, it has become feasible to compute large grids of binary evolutionary models and assess the impact of the various assumptions.  Especially from the observational side we expect large progress in the near future.  Statistics are improving with the large samples resulting from automated surveys. These will allow us to even asses the rate of IIb supernovae in different host galaxies probing different environments and even metallicity regimes.  

\section{Acknowledgements}
We thank Philip Podsiadlowski for the useful discussions regarding the comparison with his work about type IIb SNe, Matteo Cantiello  and Sung-Chul Yoon for their input about the c-IIb SNe, as well the referee for the useful feedback. 
SdM is supported by NASA through Hubble Fellowship grant HST-HF-51270.01-A awarded by the Space Telescope Science Institute, which is operated by the Association of Universities for Research in Astronomy, Inc., for NASA, under contract NAS 5-26555.

\bibliography{lit}

\Online
\onecolumn
\begin{appendix}
\section{Tables}
\begin{table*}[h!]
\caption{Properties of binary stellar models for an initial primary mass of 15~\Msun, conservative mass transfer and with variation of the initial  secondary mass M$_{\rm 2i}$ and initial orbital period P$_{\rm orb}$.}
\label{TC}
\begin{center}
\begin{tabular}{|c|c|c|c|c|c|c|c|c|c|}
\hline \hline
$N^0$ & 
\begin{tabular}{c}
$M_{2i}$ \\($M_\odot$)\\ 
\end{tabular} 
& \begin{tabular}{c}
$M_{1f}$ \\($M_\odot$)\\ 
\end{tabular} 
& \begin{tabular}{c}
$M_{2f}$ \\($M_\odot$)\\ 
\end{tabular} 
& \begin{tabular}{c} 
$P_{orb}$ \\ (Days) \\
\end{tabular} 
& \begin{tabular}{c} 
$M_{H1}$\\($M_\odot$)\\
\end{tabular} 
& log$\left(\frac{L_1}{L_\odot}\right)$ & 
\begin{tabular}{c}
$T_{\rm eff,1}$ \\($10^3$K)\\ 
\end{tabular} 
& log$\left(\frac{L_2}{L_\odot}\right)$ & 
\begin{tabular}{c}
$T_{\rm eff,2}$ \\($10^3$K)\\
\end{tabular} \\
\hline 
0a & 14.99 & 5.20 & 22.88 &1000 &  0.0997 & 5.04 & 3.75 & 5.17 & 12.7\\
0b & 14.95 & 5.19 & 22.85 & 1000 & 0.0992 & 5.04 & 3.76 & 5.17 & 13.6\\
0c & 14.9 & 5.19 & 22.80 &1000 & 0.0982 & 5.04 & 3.76 & 5.16 & 13.8\\
0d & 14.87 & 5.19 & 22.78 & 1000 & 0.0982 & 5.04 & 3.76 & 5.16 & 15.3\\
0e & 14.86 & 5.19 & 22.77 & 1000 & 0.0981 & 5.04 & 3.76 & 5.15 & 19.3\\
1a & 14.99 & 5.32 & 22.72 &1100 &  0.1283 & 5.05 & 3.69 & 5.16 & 12.4\\
1b & 14.95 & 5.32 & 22.68 & 1100 & 0.1276 & 5.05 & 3.69 & 5.16 & 13.2\\
1c & 14.9 & 5.31 & 22.64 &1100 & 0.1267 & 5.05 & 3.69 & 5.15 & 13.3\\
1d & 14.87 & 5.31 & 22.61 & 1100 & 0.1262 & 5.05 & 3.69 & 5.15 & 14.4\\
1e & 14.86 & 5.31 & 22.60 & 1100 & 0.1260 & 5.05 & 3.69 & 5.15 & 15.9\\
2a & 14.93 & 5.46 & 22.42 & 1200 & 0.1570 & 5.06 & 3.66 & 5.15 & 13.4\\
2b & 14.92 & 5.46 & 22.41 & 1200 & 0.1470 & 5.06 & 3.66 & 5.15 & 13.6\\
2c & 14.9 & 5.46 & 22.40 & 1200 & 0.1556 & 5.06 & 3.66 & 5.15 & 13.3\\
2d & 14.85 & 5.45 & 22.35 & 1200 & 0.1555 & 5.06 & 3.67 & 5.14 & 13.7\\
2e & 14.84 & 5.45 & 22.34 & 1200 & 0.1553 & 5.06 & 3.67 & 5.14 & 15.4\\
3a & 14.73 & 5.62 & 22.00 & 1300 & 0.1949 & 5.07 & 3.60 & 5.12 & 16.3\\
3b & 14.72 & 5.62 & 22.00 & 1300 & 0.1958 & 5.07 & 3.60 & 5.12 & 16.0\\
3c & 14.71 & 5.61 & 21.99 & 1300 & 0.1911 & 5.07 & 3.60 & 5.12 & 14.9\\
3d & 14.7 & 5.61 & 21.98 & 1300 & 0.1911 & 5.07 & 3.60 & 5.12 & 14.9\\
3e & 14.68 & 5.61 & 21.96 & 1300 & 0.1912 & 5.07 & 3.60 & 5.11 & 15.2\\
3f & 14.66 & 5.61 & 21.94 & 1300 & 0.2028 & 5.07 & 3.60 & 5.10 & 18.4\\
4a & 14.54 & 5.79 & 21.63 & 1400 & 0.2774 & 5.07 & 3.53 & 5.09 & 17.6\\
4b & 14.53 & 5.79 & 21.62 & 1400 & 0.2787 & 5.07 & 3.53 & 5.09 & 16.6\\
4c & 14.5 & 5.79 & 21.59 & 1400 & 0.2711 & 5.07 & 3.53 & 5.08 & 17.1\\
5a & 14.36 & 5.98 & 21.23 & 1500 & 0.3677 & 5.07 & 3.47 & 5.06 & 18.1\\
5b & 14.35 & 5.98 & 21.23 & 1500 & 0.3699 & 5.07 & 3.47 & 5.06 & 17.9\\
5c & 14.33 & 5.98 & 21.21 & 1500 & 0.3625 & 5.07 & 3.47 & 5.06 & 18.2\\
5d & 14.32 & 5.97 & 21.20 & 1500 & 0.3765 & 5.07 & 3.47 & 5.04 & 19.7\\
6a & 14.17 & 6.20 & 20.80 & 1600 & 0.4772 & 5.07 & 3.42 & 5.03 & 18.8\\
6b & 14.15 & 6.19 & 20.79 & 1600 & 0.4749 & 5.07 & 3.43 & 5.02 & 19.1\\
7a & 14.01 & 6.44 & 20.37 & 1700 & 0.6257 & 5.07 & 3.39 & 5.00 & 19.1\\
7b & 14 & 6.44 & 20.37 & 1700 & 0.6458 & 5.07 & 3.39 & 5.00 & 19.0\\
7c & 13.97 & 6.43 & 20.34 & 1700 & 0.6399 & 5.07 & 3.39 & 4.99 & 20.0\\
\hline
\end{tabular}
\tablefoot{The secondary is, for every model, at the start of mass transfer in its Hertszprung gap. The final mass of the primary and secondary, M$_{\rm f1}$ and M$_{\rm f2}$, and the amount of hydrogen in the envelope M$_{\rm H}$ are given at the time of explosion, together with the luminosity log(L/L$_\odot$) and effective temperature T$_{\rm eff}$ of the primary and secondary at the onset of central carbon burning.}
\end{center}
\end{table*}

\begin{table*}
\caption{Properties of binary stellar models for an initial primary mass of 15~\Msun, conservative mass transfer and with variation of the initial  secondary mass M$_{\rm 2i}$ and initial orbital period P$_{\rm orb}$.}
\label{HTC}
\begin{center}
\begin{tabular}{|c|c|c|c|c|c|c|c|c|c|}
\hline \hline
$N^0$ & 
\begin{tabular}{c}
$M_{2i}$ \\($M_\odot$)\\ 
\end{tabular} 
& \begin{tabular}{c}
$M_{1f}$ \\($M_\odot$)\\ 
\end{tabular} 
& \begin{tabular}{c}
$M_{2f}$ \\($M_\odot$)\\ 
\end{tabular} 
& \begin{tabular}{c} 
$P_{orb}$ \\ (Days) \\
\end{tabular} 
& \begin{tabular}{c} 
$M_{H1}$\\($M_\odot$)\\
\end{tabular} 
& log$\left(\frac{L_1}{L_\odot}\right)$ & 
\begin{tabular}{c}
$T_{\rm eff,1}$ \\($10^3$K)\\
\end{tabular} 
& log$\left(\frac{L_2}{L_\odot}\right)$ & 
\begin{tabular}{c}
$T_{\rm eff,2}$ \\($10^3$K)\\
\end{tabular} \\
	\hline 
h0a & 13.5 & 5.10 & 21.52 & 1000 &  0.0835 & 5.03 & 3.86 & 4.96 & 32.2\\
h0b & 14 & 5.13 & 21.98 & 1000 & 0.0886 & 5.03 & 3.82 & 5.01 & 31.7\\
h0c & 14.85 & 5.19 & 22.76 & 1000 & 0.0980 & 5.04 & 3.76 & 5.09 & 30.3\\
h1a & 13.3 & 5.16 & 21.25 & 1100 &  0.0997 & 5.03 & 3.76 & 4.94 & 32.3\\
h1b & 13.4 & 5.17 & 21.34 & 1100 &  0.1014 & 5.04 & 3.76 & 4.95 & 32.2\\
h1c & 14 & 5.23 & 21.85 & 1100 & 0.1111 & 5.04 & 3.73 & 5.00 & 31.7\\
h1d & 14.85 & 5.31 & 22.59 & 1100 & 0.1258 & 5.05 & 3.69 & 5.08 & 30.2\\
h2a & 13.2 & 5.31 & 20.86 & 1200 & 0.0983 & 5.05 & 3.78 & 4.92 & 32.2\\
h2b & 13.3 & 5.32 & 20.95 & 1200 & 0.1012 & 5.04 & 3.77 & 4.93 & 32.1\\
h2c & 14 & 5.37 & 21.60 & 1200 & 0.1195 & 5.06 & 3.72 & 4.99 & 31.5\\
h2d & 14.8 & 5.45 & 22.31 & 1200 & 0.1450 & 5.06 & 3.67 & 5.07 & 30.1\\
h2e & 14.83 & 5.45 & 22.33 & 1200 & 0.1450 & 5.06 & 3.67 & 5.07 & 30.1\\
h3a & 12.1 & 5.35 & 19.65 & 1300 & 0.0983 & 5.06 & 3.78 & 4.80 & 32.6\\
h3b & 12.2 & 5.36 & 19.75 & 1300 & 0.1079 & 5.06 & 3.77 & 4.81 & 32.5\\
h3c & 14.64 & 5.61 & 21.93 & 1300 & 0.2009 & 5.07 & 3.60 & 5.04 & 30.5\\
h4a & 11.4 & CONTACT & / & 1400 & / & / & / & / & /\\
h4b & 11.5 & 5.43 & 18.85 & 1400 & 0.1223 & 5.06 & 3.72 & 4.74 & 32.5\\
h4c & 11.9 & 5.47 & 19.29 & 1400 & 0.1329 & 5.06 & 3.69 & 4.77 & 32.6\\
h4d & 12 & 5.48 & 19.38 & 1400 & 0.1430 & 5.07 & 3.68 & 4.78 & 32.6\\
h4e & 14.49 & 5.79 & 21.58 & 1400 & 0.2785 & 5.07 & 3.53 & 5.01 & 30.8\\
h5a & 11.1 & CONTACT & / & 1500 & / & / & / & / & /\\
h5b & 11.2 & 5.56 & 18.37 & 1500 & 0.1615 & 5.07 & 3.65 & 4.70 & 32.4\\
h5c & 11.6 & 5.60 & 18.84 & 1500 & 0.1843 & 5.07 & 3.62 & 4.73 & 32.5\\
h5d & 13 & 5.79 & 20.06 & 1500 & 0.2711 & 5.07 & 3.53 & 4.86 & 32.3\\
h5e & 14 & 5.93 & 20.92 & 1500 & 0.3538 & 5.07 & 3.48 & 4.95 & 31.4\\
h5f & 14.31 & 5.97 & 21.19 & 1500 & 0.3604 & 5.07 & 3.47 & 4.98 & 31.0\\
h6a & 10.7 & CONTACT & / & 1600 & / & / & / & / & /\\
h6b & 10.8 & 5.68 & 17.81 & 1600 & 0.2183 & 5.07 & 3.58 & 4.64 & 32.4\\
h6c & 11.3 & 5.76 & 18.36 & 1600 & 0.2442 & 5.07 & 3.55 & 4.69 & 32.5\\
h6d & 13 & 6.01 & 19.81 & 1600 & 0.3916 & 5.07 & 3.46 & 4.84 & 32.2\\
h6e & 14.1 & 6.19 & 20.73 & 1600 & 0.4740 & 5.07 & 3.43 & 4.94 & 31.2\\
h6f & 14.14 & 6.19 & 20.77 & 1600 & 0.4862 & 5.07 & 3.43 & 4.95 & 31.1\\
h7a & 10.4 & CONTACT & / & 1700 & / & / & / & / & /\\
h7b & 10.5 & 5.81 & 17.34 & 1700 & 0.2716 & 5.08 & 3.53 & 4.61 & 32.1\\
h7c & 11.1 & 5.91 & 17.98 & 1700 & 0.3318 & 5.08 & 3.50 & 4.66 & 32.3\\
h7d & 11.4 & 5.96 & 18.23 & 1700 & 0.3468 & 5.08 & 3.48 & 4.68 & 32.4\\
h7e & 12.7 & 6.20 & 19.31 & 1700 & 0.4784 & 5.07 & 3.42 & 4.80 & 32.2\\
h7f & 12.8 & 6.21 & 19.38 & 1700 & 0.5041 & 5.07 & 3.42 & 4.81 & 32.2\\
h8a & 10.3 & CONTACT & / & 1800 & / & / & / & / & /\\
h8b & 10.4 & 5.95 & 17.08 & 1800 & 0.3467 & 5.08 & 3.49 & 4.58 & 32.0\\
h8c & 11 & 6.05 & 17.70 & 1800 & 0.4006 & 5.08 & 3.45 & 4.63 & 32.4\\
h8d & 11.5 & 6.16 & 18.11 & 1800 & 0.4668 & 5.08 & 3.43 & 4.68 & 32.2\\
h8e & 11.8 & 6.22 & 18.34 & 1800 & 0.4860 & 5.08 & 3.42 & 4.70 & 32.2\\
h8f & 11.9 & 6.24 & 18.41 & 1800 & 0.5104 & 5.08 & 3.42 & 4.71 & 32.2\\
h9a & 10.3 & CONTACT & / & 1900 & / & / & / & / & /\\
h9b & 10.4 & 6.12 & 16.95 & 1900 & 0.4531 & 5.08 & 3.44 & 4.57 & 31.8\\
h9c & 10.8 & 6.19 & 17.24 & 1900 & 0.4873 & 5.08 & 3.42 & 4.60 & 28.8\\
h9d & 10.9 & 6.22 & 17.31 & 1900 & 0.5014 & 5.08 & 3.42 & 4.61 & 32.0\\
\hline 
\end{tabular}
\tablefoot{The secondary is, for every model, at the start of mass transfer a main sequence star. The other symbols have a similar meaning as in Table \ref{TC}. If the evolution of the binary system ended because of formation of a contact binary, the parameters cannot be determined and are indicated by '/'.}
\end{center}
\end{table*}

\begin{table*}
\caption{Properties of binary stellar models for an initial primary mass of 15~\Msun, conservative mass transfer and with variation of the initial  secondary mass M$_{\rm 2i}$ and initial orbital period P$_{\rm orb}$.}
\label{GTC}
\begin{center}
\begin{tabular}{|c|c|c|c|c|c|c|c|c|c|c|c|}
\hline \hline
$N^0$ & 
\begin{tabular}{c}
$M_{2i}$\\($M_\odot$)\\ 
\end{tabular} 
& \begin{tabular}{c}
$M_{1f}$\\($M_\odot$)\\ 
\end{tabular} 
& \begin{tabular}{c}
$M_{2f}$\\($M_\odot$)\\ 
\end{tabular} 
& \begin{tabular}{c} 
$P_{orb}$\\(Days)\\
\end{tabular} 
& \begin{tabular}{c} 
$M_{H1}$\\($M_\odot$)\\
\end{tabular} 
& log$\left(\frac{L_1}{L_\odot}\right)$ & 
\begin{tabular}{c}
$T_{\rm eff,1}$ \\($10^3$K)\\
\end{tabular} 
& log$\left(\frac{L_2}{L_\odot}\right)$ & 
\begin{tabular}{c}
$T_{\rm eff,2}$ \\($10^3$K)\\
\end{tabular}
& $*1/*2$ &
\begin{tabular}{c}
$\not=$ time\\(yr)\\ 
\end{tabular} \\
	\hline  
g2a & 14.94 & 5.60 & 22.26 & 1200 & 0.2178 & 5.06 & 3.66 & 5.08 & 3.63 & *2 & 14200\\
g2b & 14.99 & 5.61 & 21.75 & 1200 & 0.2211 & 5.06 & 3.66 & 5.08 & 3.63 & *2 & 70400\\
g3a & 14.74 & 5.62 & 22.01 & 1300 & 0.2061 & 5.07 & 3.60 & 4.90 & 3.79 & *1 & /\\
g3b & 14.75 & 5.62 & 22.02 & 1300 & 0.2058 & 5.07 & 3.60 & 4.90 & 3.79 & *1 & /\\
g3c & 14.93 & 5.64 & 22.18 & 1300 & 0.2036 & 5.07 & 3.59 & 5.02 & 3.67 & *1 & /\\
g3d & 14.94 & 5.76 & 22.18 & 1300 & 0.2638 & 5.07 & 3.59 & 5.07 & 3.64 & *2 & 300\\
g3e & 14.99 & 5.80 & 21.37 & 1300 & 0.2894 & 5.07 & 3.58 & 5.08 & 3.62 & *2 & 63100\\
g4a & 14.55 & 5.79 & 21.64 & 1400 & 0.2761 & 5.07 & 3.53 & 4.86 & 3.83 & *1 & /\\
g4b & 14.9 & 5.84 & 21.94 & 1400 & 0.3081 & 5.07 & 3.51 & 4.95 & 3.74 & *1 & /\\
g4c & 14.94 & 5.84 & 21.98 & 1400 & 0.3087 & 5.07 & 3.51 & 5.04 & 3.66 & *1 & /\\
g4d & 14.95 & 6.02 & 21.79 & 1400 & 0.4012 & 5.07 & 3.51 & 5.07 & 3.63 & *2 & 7900\\
g4e & 14.99 & 6.03 & 21.01 & 1400 & 0.4012 & 5.07 & 3.51 & 5.08 & 3.61 & *2 & 58300\\
g5a & 14.37 & 5.98 & 21.24 & 1500 & 0.3656 & 5.07 & 3.47 & 4.81 & 3.85 & *1 & /\\
g5b & 14.4 & 5.99 & 21.27 & 1500 & 0.3728 & 5.07 & 3.47 & 4.82 & 3.85 & *1 & /\\
g5c & 14.55 & 6.01 & 21.40 & 1500 & 0.3904 & 5.07 & 3.45 & 4.85 & 3.82 & *1 & /\\
g5d & 14.9 & 6.06 & 21.69 & 1500 & 0.4240 & 5.07 & 3.45 & 4.95 & 3.74 & *1 & /\\
g5e & 14.94 & 6.06 & 21.73 & 1500 & 0.4244 & 5.07 & 3.45 & 5.04 & 3.66 & *1 & /\\
g5f & 14.95 & 6.27 & 21.56 & 1500 & 0.5305 & 5.07 & 3.45 & 5.07 & 3.63 & *2 & 9100\\
g5g & 14.99 & 6.28 & 20.56 & 1500 & 0.5328 & 5.07 & 3.45 & 5.07 & 3.60 & *2 & 59500\\
g6a & 14.18 & 6.20 & 20.81 & 1600 & 0.4796 & 5.07 & 3.42 & 4.78 & 3.88 & *1 & /\\
g6b & 14.2 & 6.20 & 20.83 & 1600 & 0.4980 & 5.07 & 3.42 & 4.77 & 3.88 & *1 & /\\
g6c & 14.3 & 6.22 & 20.91 & 1600 & 0.4997 & 5.07 & 3.42 & 4.80 & 3.86 & *1 & /\\
g6d & 14.4 & 6.23 & 20.99 & 1600 & 0.5236 & 5.07 & 3.42 & 4.82 & 3.85 & *1 & /\\
\hline
\end{tabular}
\tablefoot{The secondary is, for every model, at the start of mass transfer on the giant branch. The column with *1/*2 indicates which star in the binary system explodes first. The time difference between the two stars reaching carbon burning is indicated if star 2 explodes before star 1. The other symbols have a similar meaning as in Table \ref{TC}.}
\end{center}
\end{table*}

\label{lastpage}

\begin{table*}
\footnotesize
\caption{Properties of binary stellar models for an initial primary mass of 15~\Msun, with variation of the initial  secondary mass M$_{\rm 2i}$, variation of the accretion efficiency $\beta$ and initial orbital period P$_{\rm orb}$.} 
\label{TNC}
\begin{center}
\begin{tabular}{|c|c|c|c|c|c|c|c|c|c|c|c|c|}
\hline \hline 
$N^0$ & 
\begin{tabular}{c}
$M_{2i}$\\($M_\odot$)\\ 
\end{tabular} 
& \begin{tabular}{c}
$M_{1f}$\\($M_\odot$)\\ 
\end{tabular} 
& \begin{tabular}{c}
$M_{2f}$\\($M_\odot$)\\ 
\end{tabular} 
& \begin{tabular}{c} 
$P_{orb}$\\(Days)\\
\end{tabular} 
& $\beta$
& \begin{tabular}{c} 
$M_{H1}$\\($M_\odot$)\\
\end{tabular} 
& log$\left(\frac{L_1}{L_\odot}\right)$ & 
\begin{tabular}{c}
$T_{\rm eff,1}$ \\($10^3$K)\\
\end{tabular} 
& log$\left(\frac{L_2}{L_\odot}\right)$ & 
\begin{tabular}{c}
$T_{\rm eff,2}$ \\($10^3$K)\\
\end{tabular}
& $*1/*2$ &
\begin{tabular}{c}
$\not=$ time\\(yr)\\ 
\end{tabular} \\
	\hline 
8.5 & 14.9 & 5.15 & 18.86 & 800 & 0.50 & 0.0897 & 5.04 & 3.84 & 5.08 & 3.56 & *1 & /\\
9.9 & 14.9 & 5.13 & 22.07 & 900 & 0.90 & 0.0814 & 5.03 & 3.84 & 5.14 & 12.6 & *1 & /\\
9.5 & 14.9 & 5.38 & 18.75 & 900 & 0.50 & 0.1671 & 5.04 & 3.71 & 5.10 & 3.55 & *2 & 11900\\
0c9 & 14.9 & 5.23 & 21.97 & 1000 & 0.90 & 0.1050 & 5.04 & 3.73 & 5.13 & 11.8 & *1 & /\\
1b9 & 14.95 & 5.50 & 21.41 & 1100 & 0.90 & 0.2107 & 5.05 & 3.66 & 5.08 & 3.62 & *2 & 20200\\
1b7 & 14.95 & 5.60 & 19.93 & 1100 & 0.70 & 0.2469 & 5.05 & 3.62 & 5.07 & 3.59 & *2 & 16700\\
1c9 & 14.9 & 5.36 & 21.81 & 1100 & 0.90 & 0.1512 & 5.05 & 3.66 & 5.12 & 10.7 & *1 & /\\
1c5 & 14.9 & 5.73 & 18.54 & 1100 & 0.50 & 0.2878 & 5.06 & 3.58 & 5.09 & 3.55 & *2 & 12300\\
1d75 & 14.87 & 5.43 & 20.56 & 1100 & 0.75 & 0.1712 & 5.05 & 3.63 & 5.08 & 8.81 & *1 & /\\
1d5 & 14.87 & 5.57 & 18.57 & 1100 & 0.50 & 0.2680 & 5.06 & 3.58 & 4.94 & 3.69 & *1 & /\\
1d25 & 14.87 & 5.74 & 16.67 & 1100 & 0.25 & 0.2857 & 5.06 & 3.53 & 4.93 & 3.64 & *1 & /\\
2c5 & 14.9 & 5.76 & 18.47 & 1200 & 0.50 & 0.2661 & 5.07 & 3.54 & 4.95 & 3.66 & *1 & /\\
3d5 & 14.7 & 5.96 & 18.16 & 1300 & 0.50 & 0.3658 & 5.07 & 3.48 & 4.89 & 3.71 & *1 & /\\
4a7 & 14.54 & 6.01 & 19.34 & 1400 & 0.70 & 0.3906 & 5.07 & 3.46 & 4.85 & 3.77 & *1 & /\\
4c9 & 14.5 & 5.96 & 19.68 & 1400 & 0.90 & 0.3671 & 5.07 & 3.47 & 5.02 & 12.7 & *1 & /\\
4c7 & 14.5 & 6.00 & 19.31 & 1400 & 0.70 & 0.3873 & 5.07 & 3.46 & 5.00 & 9.64 & *1 & /\\
4c5 & 14.5 & 6.18 & 17.84 & 1400 & 0.50 & 0.4844 & 5.07 & 3.43 & 4.86 & 3.74 & *1 & /\\
g4b7 & 14.9 & 6.06 & 19.66 & 1400 & 0.70 & 0.4204 & 5.07 & 3.45 & 4.95 & 3.69 & *1 & /\\
g4c7 & 14.94 & 6.07 & 19.70 & 1400 & 0.70 & 0.4191 & 5.07 & 3.45 & 5.04 & 3.61 & *1 & /\\
g4d7 & 14.95 & 6.28 & 19.60 & 1400 & 0.70 & 0.5387 & 5.07 & 3.45 & 5.07 & 3.58 & *2 & 9700\\
h5.0 & 8.3 & 5.37 & 10.15 & 1500 & 0.25 & 0.0795 & 5.07 & 3.87 & 3.88 & 25.1 & *1 & /\\
h5.1 & 8.6 & 5.41 & 10.44 & 1500 & 0.25 & 0.0912 & 5.07 & 3.79 & 3.93 & 25.4 & *1 & /\\
h5.2 & 8.8 & 5.44 & 10.64 & 1500 & 0.25 & 0.1074 & 5.07 & 3.75 & 3.96 & 25.5 & *1 & /\\
h5.3 & 8.8 & 5.41 & 12.49 & 1500 & 0.50 & 0.0961 & 5.07 & 3.80 & 4.17 & 27.8 & *1 & /\\
h5.4 & 9.1 & 5.44 & 12.78 & 1500 & 0.50 & 0.1164 & 5.07 & 3.75 & 4.20 & 28.0 & *1 & /\\
h5.5 & 9.4 & 5.55 & 11.22 & 1500 & 0.25 & 0.1510 & 5.07 & 3.65 & 4.06 & 26.0 & *1 & /\\
h5.6 & 9.4 & 5.49 & 13.07 & 1500 & 0.50 & 0.1239 & 5.07 & 3.70 & 4.24 & 28.2 & *1 & /\\
h5.7 & 9.4 & CONTACT & / & 1500 & 0.75 & / & / & / & / & / & / & /\\
h5.8 & 9.7 & CONTACT & / & 1500 & 0.75 & / & / & / & / & / & / & /\\
h5.9 & 10 & 5.50 & 15.49 & 1500 & 0.75 & 0.1308 & 5.07 & 3.69 & 4.46 & 30.5 & *1 & /\\
h5d5 & 14 & 6.38 & 17.23 & 1500 & 0.50 & 0.5975 & 5.07 & 3.40 & 4.43 & 27.1 & *1 & /\\
5b9 & 14.35 & 6.06 & 20.46 & 1500 & 0.90 & 0.4163 & 5.07 & 3.45 & 5.04 & 16.6 & *1 & /\\
5b8 & 14.35 & 6.14 & 19.72 & 1500 & 0.80 & 0.4624 & 5.07 & 3.43 & 5.01 & 14.3 & *1 & /\\
5b75 & 14.35 & 6.19 & 19.35 & 1500 & 0.75 & 0.4886 & 5.07 & 3.43 & 4.97 & 8.67 & *1 & /\\
5b7 & 14.35 & 6.24 & 18.97 & 1500 & 0.70 & 0.5176 & 5.07 & 3.42 & 4.84 & 3.77 & *1 & /\\
5b6 & 14.35 & 6.34 & 18.25 & 1500 & 0.60 & 0.5750 & 5.07 & 3.40 & 4.81 & 3.78 & *1 & /\\
5b5 & 14.35 & 6.45 & 17.54 & 1500 & 0.50 & 0.6325 & 5.07 & 3.39 & 4.81 & 3.77 & *1 & /\\
5b25 & 14.35 & 6.74 & 15.87 & 1500 & 0.25 & 0.8062 & 5.07 & 3.37 & 4.80 & 3.72 & *1 & /\\
g5c5 & 14.55 & 6.49 & 17.72 & 1500 & 0.50 & 0.6595 & 5.07 & 3.39 & 4.85 & 3.79 & *1 & /\\
6b9 & 14.15 & 6.29 & 20.04 & 1600 & 0.90 & 0.5445 & 5.07 & 3.41 & 5.00 & 17.5 & *1 & /\\
\hline
\end{tabular}
\tablefoot{The column with *1/*2 indicates which star in the binary system explodes first. The time difference between the two stars reaching carbon burning is indicated if star 2 explodes before star 1. If the evolution of the binary system ended because of formation of a contact binary, the parameters cannot be determined and are indicated by '/'. The other symbols have a similar meaning as in Table \ref{TC}.}
\end{center}
\end{table*}
\end{appendix}

\end{document}